\documentclass[11pt]{article}
\usepackage[margin=1in]{geometry}
\usepackage{times}
\usepackage{amsmath,amssymb}
\usepackage{graphicx}
\usepackage{natbib}
\usepackage{hyperref}

\PassOptionsToPackage{numbers, compress}{natbib}
\usepackage{enumitem}
\usepackage{listings}
\usepackage{amsmath}
\usepackage{epigraph}
\usepackage[utf8]{inputenc} 
\usepackage[T1]{fontenc}    
\usepackage{hyperref}       
\usepackage{url}            
\usepackage{booktabs}       
\usepackage{amsfonts}       
\usepackage{nicefrac}       
\usepackage{microtype}      
\usepackage{xcolor}         
\usepackage{graphicx}

\usepackage{wrapfig}
\usepackage[framemethod=TikZ]{mdframed}
\mdfsetup{%
    middlelinecolor =   none,
    middlelinewidth =   1pt,
    backgroundcolor =   blue!5,
    roundcorner     =   5pt,
}
\newmdenv[
    middlelinecolor=none,
    middlelinewidth=1pt,
    backgroundcolor=blue!5,
    roundcorner=5pt
]{bluebox}

\newmdenv[
    middlelinecolor=none,
    middlelinewidth=1pt,
    backgroundcolor=gray!20,
    roundcorner=5pt
]{graybox}
\usepackage{xcolor}
\usepackage{tcolorbox}
\newtcbox{\grayboxtext}[1][]{
    on line,
    colframe=gray!20,
    colback=gray!20,
    boxrule=0.5pt,
    arc=4pt,
    boxsep=0pt,
    left=2pt,
    right=2pt,
    top=1pt,
    bottom=1pt,
    #1
}
\newtcbox{\blueboxtext}[1][]{
    on line,
    colframe=blue!5,
    colback=blue!5,
    boxrule=0.5pt,
    arc=4pt,
    boxsep=0pt,
    left=2pt,
    right=2pt,
    top=1pt,
    bottom=1pt,
    #1
}
\title{OpenReview Should be Protected and Leveraged as a \\ Community Asset for Research in the Era of Large Language Models}

\author{%
  Hao Sun \\
  University of Cambridge\\
  \texttt{hs789@cam.ac.uk} \\
  \and
  Yunyi Shen \\
  MIT \\
  \texttt{yshen99@mit.edu} \\
  \and
  Mihaela van der Schaar \\
  University of Cambridge \\
  \texttt{mv472@cam.ac.uk} \\
}

\begin{document}

\maketitle

\begin{abstract}
  In the era of large language models (LLMs), high-quality, domain-rich, and continuously evolving datasets capturing expert-level knowledge, core human values, and reasoning are increasingly valuable. \textbf{This position paper argues that OpenReview --- the continually evolving repository of research papers, peer reviews, author rebuttals, meta-reviews, and decision outcomes --- should be leveraged more broadly as a core \textit{community asset} for advancing research in the era of LLMs.} We highlight three promising areas in which OpenReview can uniquely contribute: enhancing the quality, scalability, and accountability of peer review processes; enabling meaningful, open-ended benchmarks rooted in genuine expert deliberation; and supporting alignment research through real-world interactions reflecting expert assessment, intentions, and scientific values. To better realize these opportunities, we suggest the community collaboratively explore standardized benchmarks and usage guidelines around OpenReview, inviting broader dialogue on responsible data use, ethical considerations, and collective stewardship.

\end{abstract}

\section{Introduction}
\begin{quote}
\itshape
"\textbf{Knowledge belongs to humanity, and is the torch which illuminates the world.}"

\hfill --- Louis Pasteur
\end{quote}
The past years have witnessed an extraordinary shift in the role of data within machine learning~\citep{zha2025data,seedat2023navigating}, especially with the recent advances of large language models (LLMs)~\citep{brown2020language,chowdhery2022palm,touvron2023llama}, which have progressed from task-specific tools to general-purpose reasoning engines~\citep{devlin2019bert,jaech2024openai,openai2025deepresearch}. As their capabilities expand across domains, the role of data for training, evaluation, and alignment becomes even more important~\citep{zhou2023lima,gadre2023datacomp,kaplan2020scaling,hoffmann2022training}. The current wave of LLM development increasingly depends on high-quality, human-centered feedback~\citep{christiano2017deep,ouyang2022training,bai2022training,stiennon2020learning,silver2025welcome}, not only for fine-tuning and instruction adherence, but also for assessing model behavior, identifying failure modes, and aligning outputs with human expectations~\citep{chiang2024chatbot,farquhar2024detecting,lindsey2025biology,lambert2024rewardbench}. Yet many of the datasets used for these purposes remain limited in coverage~\citep{bai2022constitutional}, synthetic in composition~\citep{dubois2024alpacafarm,gao2023scaling}, or static in structure~\citep{bai2022training}. As a result, they often fail to capture the complexity, disagreement, and subtle reasoning that characterize authentic human judgment~\citep{kahneman2021noise,wang2023openchat,liu2024aligning,liu2025aligning}.

At the same time, the powerful capabilities of LLMs are beginning to reshape scientific workflows themselves~\citep{karabacak2023embracing,ganjavi2024publishers,Elsevier2023_ScopusAI,van2023chatgpt,fecher2025friend}. Tools based on LLMs such as ChatGPT are making research communication, including literature reviews and even paper writing, more accessible~\citep{Taylor2022_Galactica,else2023chatgpt,si2024can,wang2024autosurvey}, hence accelerating scientific output and contributing to a significant rise in the volume of submissions to top conferences. Such a transformation has intensified pressure on the peer review system~\citep{Horbach2019_PeerReviewPressure,AX2025-doge}.
Conferences now receive more than 10 thousands of submissions per cycle, and the human effort required to maintain high-quality, fair, and constructive reviewing has become difficult to sustain. Given such high pressure, the need for scalable assistance tools, better evaluation data, and models that can understand or generate scholarly critique has grown~\citep{AX2025-doge,chiang2023can,kumar2024towards}. However, large-scale, systematic exploration regarding both the datasets and methodologies that enable LLMs to capture the richness of peer review interactions is still missing~\citep{liu2023reviewergpt,zhou2024llm,thakkar2025can}.

OpenReview\footnote{\url{https://openreview.net/}}\citep{soergel2013open}, the public review platform widely used by conferences such as ICLR, NeurIPS, and others, offers a unique opportunity to meet the needs of both sides. Contributed by the community and continually expanding over time, OpenReview hosts large-scale, structured records of scientific discussion, typically including paper submissions, reviewer assessments, author rebuttals, meta-reviews, and final decisions. These interactions span multiple rounds and involve diverse expert perspectives, making OpenReview an invaluable living dataset grounded in real-world scientific research deliberation. And has the potential to enrich both data-centric LLM research and assist the peer review system.

\begin{figure}[t!]
    \centering
    \includegraphics[width=1.0\linewidth]{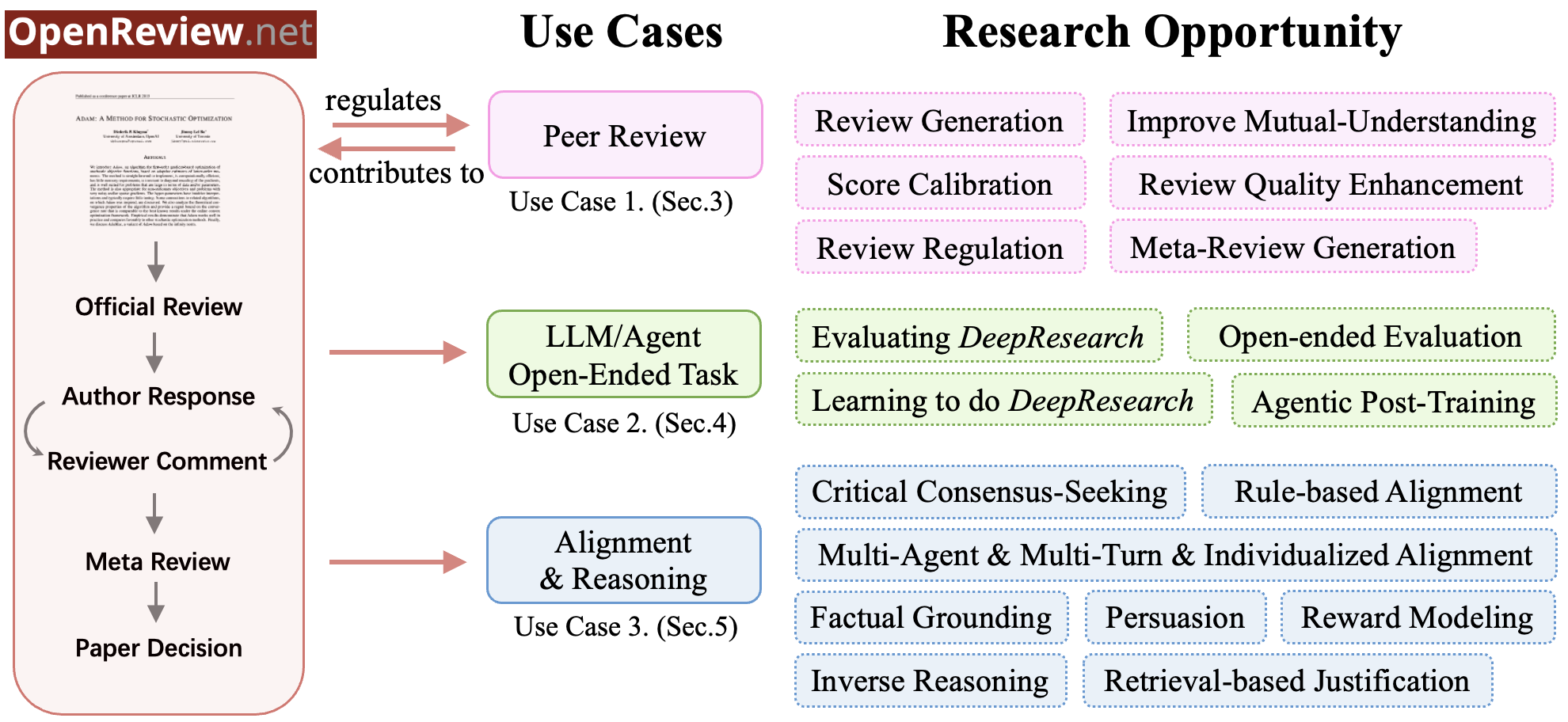}\vspace{-0.1cm}
    \caption{\small \textbf{Left}: an overview of the OpenReview data generation process; \textbf{mid}: this position paper argues OpenReview supports three main valuable applications --- regulating peer review, empowering LLM and Agentic open-ended task research, and post-training for alignment and reasoning; \textbf{right}: highlighted research opportunities around those use cases.}
    \label{fig:teaser_fig}\vspace{-0.4cm}
\end{figure}

\textbf{This position paper argues that OpenReview should be leveraged more broadly as a core community asset for advancing research in the era of LLMs.} We elaborate on three areas where this dataset can provide immediate value: 
\begin{enumerate}[leftmargin=*]
    \item \textbf{A data-driven approach to improve the quality and scalability of peer review.} OpenReview provides a unique source of structured, expert-generated assessments that can be used to train machine learning models to analyze and support the peer review process. Machine learning models, including the state-of-the-art general purpose language models~\citep{gemini25pro_2025,gpt45_2025,claude3_2024,grok3_2025}, may learn to assist reviewers in 
    drafting constructive feedback, calibrating scores, and identifying argumentative gaps, as well as summarizing responses, checking code, or detecting unhelpful language.
    In the face of rising submission volumes and reviewer fatigue, such tools could support more consistent, fair, and informative evaluations. Equally important, improving and regularizing the review process is a \textbf{prerequisite} for sustaining the long-term development of LLM-based systems that depend on high-quality expert feedback~\citep{shumailov2025author}.
    \item \textbf{Providing expert-generated benchmarks for LLM open-ended task evaluation and post-training.} 
    Open-ended tasks such as academic writing, research evaluation, persuasion, or summarization are increasingly recognized as central to the development of general-purpose AI systems and the path toward superintelligence~\citep{hughes2024open}. However, both training and evaluating models on such tasks remain challenging due to the open-ended nature and the lack of scalable, high-quality human feedback~\citep{openai2025deepresearch}.
    To this end, OpenReview offers a unique, high-quality resource: it contains expert-curated, multi-dimensional evaluations of research contributions grounded in real-world scientific progress. Its diverse content enables the design of benchmarks for open-ended tasks such as writing~\citep{gooding2025writing}, research evaluation~\citep{chiang2023can}, persuasion~\citep{khan2024debating, rogiers2024persuasion, breum2024persuasive}, and summarization~\citep{stiennon2020learning,peerread}, providing valuable data for both open-ended LLM and agentic post-training and evaluation~\citep{sun2improving}.
    \item \textbf{Supporting multi-dimensional alignment and reasoning research through scientific writing and discussion.}
    Existing benchmarks for alignment and reasoning often rely on static, synthetic, or crowd-sourced datasets that lack the depth and nuance of real expert deliberation~\citep{bai2022training,chang2024survey,cui2023ultrafeedback,dubois2024length,zhao20251,2025synthetic1}. In contrast, OpenReview offers a setting that inherently involves alignment and reasoning through evidence-based debate, disagreement, revision, and consensus building.
    This setting enables rich evaluation tasks such as score justification via retrieval-based reasoning~\citep{guu2020retrieval,lewis2020retrieval,kojima2022large,wei2022chain} and decision prediction grounded in free-form critique~\citep{minaee2021deep}. These tasks can serve as realistic testbeds for assessing how well LLMs can interpret, reason about, and align with expert judgments in the scientific research domain.
    Moreover, the dialogic nature of OpenReview --- spanning rebuttals, conflicting views, and negotiated outcomes --- offers a unique opportunity to study value pluralism, debate-style alignment in the wild~\citep{khan2024debating,liang2023encouraging,du2023improving,sorensen2024roadmap,feng2024modular}.
\end{enumerate}
To help realize the potential in and beyond those outlined use cases, we propose initial directions for community-driven benchmark development and responsible data stewardship.
Finally, we reflect on alternative perspectives, aiming to spark productive dialogue on the challenges and risks of leveraging the OpenReview as a core community asset.
\section{The State of OpenReview Now: Scale, Opportunity, and Emerging Risks}
This section examines the OpenReview platform through three perspectives. We begin with a statistical overview of its scale and evolution, using ICLR as a case study. We then highlight its value as a unique community-curated dataset for machine learning research, before turning to the structural risks that threaten its long-term quality and integrity.

\subsection{The Scale and Structure of Conference Data on OpenReview}

OpenReview provides a centralized platform for peer review and community discussion at major machine learning conferences, including ICLR, NeurIPS, and others. It preserves structured records of submissions, reviews, rebuttals, and decisions, creating a longitudinal archive of real-world expert deliberation under consistent guidelines.

\begin{figure}[t!]
    \centering
    \includegraphics[width=1.0\linewidth]{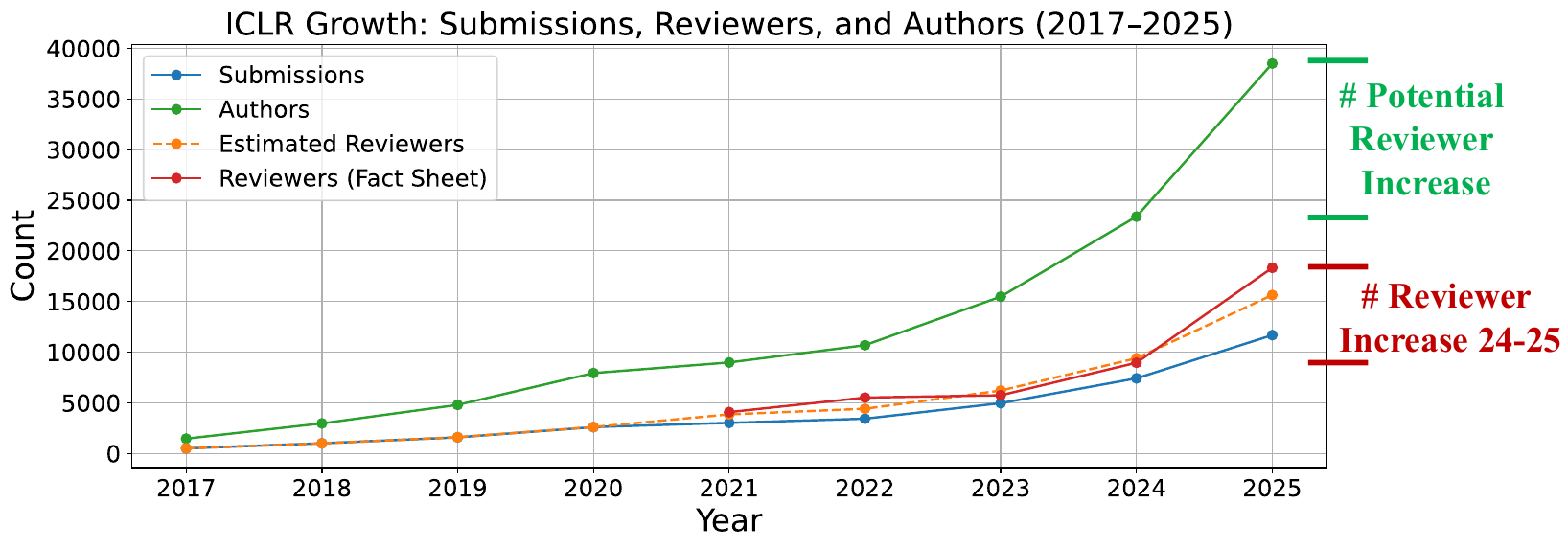} \vspace{-0.4cm}
    \caption{\small Growth trends at ICLR (2017–2025) in submissions, authors, and reviewers. While the number of reviewers has increased over time, it has not kept pace with the growth in submissions and authors, indicating a growing strain on the peer review process. The reviewer number estimation is calculated according to the number of submissions, the total number of reviews received, and the average reviewer workload of 3 per reviewer.}
    \label{fig:iclr_growth}\vspace{-0.2cm}
\end{figure}

To illustrate the scale of this platform, we focus on ICLR as a representative case. From 2017 to 2025, the number of submissions grew from fewer than 500 to over 11,600 annually. The corresponding number of authors increased from about 1,500 to 38,500, and the estimated number of reviewers rose from under 1,000 to more than 18,300. Each submission typically receives three or more expert reviews, resulting in tens of thousands of reviewer–author interactions each year. Figure~\ref{fig:iclr_growth} shows this growth trajectory in authorship, reviewing, and participation.\footnote{Data Source: ICLR 2021-2025 Fact Sheet~\citep{iclr2021factsheet,iclr2022factsheet,iclr2023factsheet,iclr2024factsheet,iclr2025factsheet}, PaperCopilot~\citep{yang2025paper}.}




\subsection{A Rapidly Growing Community Asset for Learning}

Beyond its scale, OpenReview is distinguished by its unique data quality. Unlike synthetic or crowd-sourced datasets, it captures expert-authored evaluations tied to real submissions and decisions, grounded in shared scientific norms. Each paper serves as a self-contained research scenario, typically accompanied by multiple reviews, optional rebuttals, meta-reviews, and final outcomes.

Between 2017 and 2025, ICLR alone contributed over 36,000 such interaction threads, spanning both accepted and rejected submissions. These interactions provide rich examples of open-ended scientific exploration efforts. They illustrate how researchers conduct and evaluate solutions to open questions, respond to disagreement, clarify claims, and finally construct consensus, making them highly suitable for training and evaluating LLMs on scientific reasoning, argumentation, and alignment.

Moreover, OpenReview is a continuously evolving dataset. Each year brings new topics, new papers, and new debates, reflecting both the state of research and the shifting consensus of the community. This ongoing refresh ensures its competence as a benchmark for real-world LLM deployment. In Table~\ref{tab:unified_openreview_corrected}, we compare relevant tasks in the LLM post-training community to demonstrate the general potential of the OpenReview dataset.

\begin{table}[ht]
\centering
\caption{\small Comparing datasets related to OpenReview. We will elaborate how to leverage OpenReview beyond those tasks in Sec.3-5.}
\label{tab:unified_openreview_corrected}
\resizebox{\textwidth}{!}{%
\begin{tabular}{lllccc}
\toprule
Dataset & Task & Size & Expert & Updates & OpenEnded \\
\midrule
\citet{see2017get} & Summarization & 310K & \checkmark & $\times$ & \checkmark \\
\citet{narayan2018don} & Summarization & 226K & \checkmark & $\times$ & \checkmark \\
\citet{yang2018hotpotqa} & Multi-hop QA & 113K QA pairs & $\times$ & $\times$ & \checkmark \\
\citet{rajpurkar2016squad} & Comprehension & 107K QA pairs & $\times$ & $\times$ & $\times$ \\
\citet{fan2019eli5} & Long-form QA & 270K threads & $\times$ & $\times$ & \checkmark \\
\citet{ziegler2019fine} & Preference Modeling & 60K comparisons & $\sim$ & $\times$ & \checkmark \\
\citet{bai2022training} & Alignment / Dialogue & 170K comparisons & $\sim$ & $\times$ & \checkmark \\
\citet{openassistant} & Dialogue / Alignment & 10K trees, 161K msg & $\sim$ & $\times$ & \checkmark \\
\citet{wang2019persuasion} & Argumentation & 1K dialogues & $\times$ & $\times$ & \checkmark \\
\citet{peerread} & Review, Decision & 14.7K subs, 10.7K revs & \checkmark & $\times$ & $\times$ \\
\citet{asapreview} & Aspect Rating (zh) & 46.7K reviews & $\sim$ & $\times$ & $\times$ \\
\citet{jitsupeer} & Argumentation & 2.3K & $\sim$ & $\times$ & \checkmark \\
\citet{disapere} & Argumentation & 506 threads & \checkmark & $\times$ & $\times$ \\
\citet{argscichat} & Argumentation & 41 dialogues & \checkmark & $\times$ & \checkmark \\
\midrule
OpenReview~\citep{soergel2013open} & \textbf{All above} & 36K subs, 100K+ revs & \checkmark & \checkmark & \checkmark \\
\bottomrule
\end{tabular}
}\vspace{-0.2cm}
\end{table}

\subsection{Quality Under Pressure --- The Compounding Risk of Rapid Growth}

While the growth of OpenReview presents significant opportunities, it also introduces structural risks. The rapid increase in submission volume has not been matched by a proportional increase in highly experienced reviewers. As conferences scale, an increasing fraction of reviews are written by newer or less engaged participants. This trend raises concerns about the consistency, reliability, and long-term stability of individual review signals, as well as the dataset quality and diversity~\citep{shumailov2025author}.

More precisely, the concern is not only that current reviewers may deviate from academic standards, but that a growing number of untrained reviewers may internalize and reproduce biased practices, gradually compounding the problem across generations. If evaluations are learned by imitation, biased or inconsistent norms can propagate, leading to long-term degradation of review quality. 

To formalize this concern, we provide a simplistic population genetic model to capture our intuition that a fast-growing reviewer body's lack of training can have a long-lasting effect even after the field matures by setting up precedent. Note that this is an extremely simplified model, and we acknowledge that reviewer quality is not binary and can depend on many factors.

We follow the standard Wright-Fisher model in population genetics. For each review round $t$, there are $G_t$ ``good'' reviews and $B_t$ bad reviews (in total $N_{t}=G_t+B_t$ reviews). In generation $t+1$, for $N_{t+1}$ new reviews, we model them as generated randomly, with some level of preference. Formally 
\begin{equation}
    B_{t+1}\sim \text{Binomial}\left(N_{t+1}, \frac{B_t}{(1+s(t))G_t+B_t}\right)
\end{equation}
where $s(t)$ is a factor for preference that could change over time, ideally $s(t)>0$ so that one has a preference towards writing less bad reviews than simply replicating what was seen in the past cycle. We define $X_t=\frac{B_t}{N_t}$.

We can take a diffusion limit of the model, and the proportion of bad reviews can be approximated as a Wright-Fisher SDE
\begin{equation}
    dX_t=s(t)X_t(1-X_t)dt+\sqrt{\frac{X_t(1-X_t)}{N(t)}}dW_t
\end{equation}
where $W_t$ is a one-dimensional Brownian motion.

\begin{figure}[t!]
    \centering
    \includegraphics[width=0.98\linewidth]{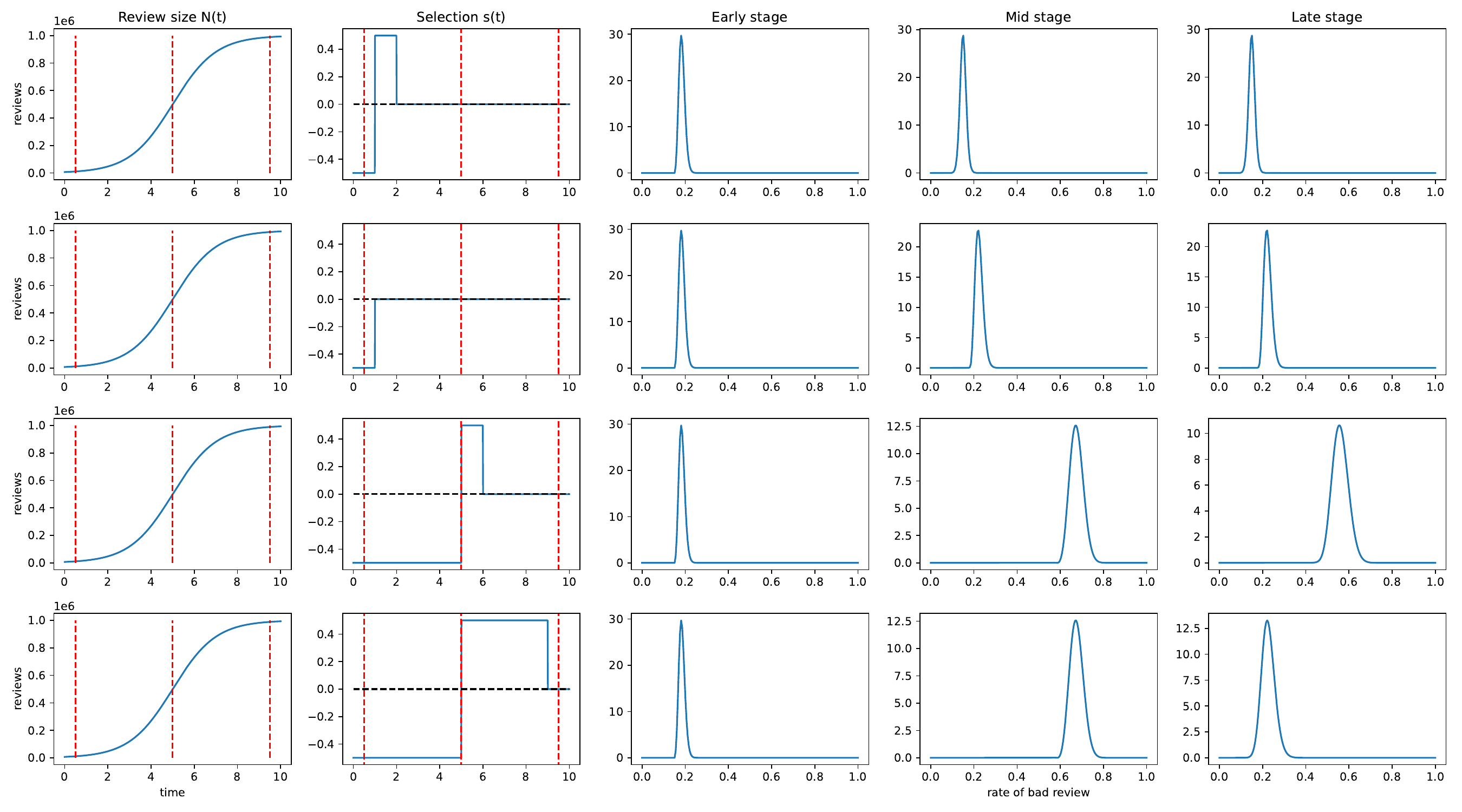}\vspace{-0.2cm}
    \caption{\small Distribution of frequency of bad reviews under Wright-Fisher type of selection model. The three stages of time are marked in red vertical lines in the first two panels. First column: model number of reviews, Second: what selection we put at which time, Third-last: distribution of proportion of bad reviews.}
    \label{fig:wrightfisher}
\end{figure}

We numerically solve the corresponding Fokker-Planck equation for different $N(t)$ and intervention $s(t)$ and present the results visualization in Figure~\ref{fig:wrightfisher}. We assume that $N(t)$ follow a logistic growth representing the usual maturing of the field. \textbf{The takeaway message is that we need to act early in stopping the trend of preferring low-quality reviews to prevent the downgrade of overall quality and set the precedent for the next generation to follow.} The trend can still be reversed in a mid to late stage, but requires more efforts (cf. first and last row in Figure~\ref{fig:wrightfisher}, we need a longer period of selection if we started late). It is useful even just to stop, instead of reverting, the current trend of preferring bad reviews (cf. second row of Figure~\ref{fig:wrightfisher}). The intuition behind these results is that if a once rare bad review was fixed into the norm during the expansion of the field, it will be part of the norm and hard to be filtered out in the future when the field grows even larger. 

\begin{mdframed}[innertopmargin=3pt,leftmargin=0pt,rightmargin=0pt,innerleftmargin=10pt,innerrightmargin=10pt,skipbelow=0pt]
\textbf{\textcolor{brown}{Take Action Now.}}
Our analysis suggests that early intervention is critical: corrective action taken before problematic patterns become institutionalized is significantly more effective than attempting to reverse them later. Proactive steps are thus essential to preserve long-term alignment between reviewing practice and community values.
\textbf{Taking action now in the early stage of the field's expansion is more effective than taking action later on when substandard review practices become the norm.}
\end{mdframed}

For OpenReview to remain a robust and trustworthy resource, its quality must be actively protected. This includes better reviewer recruitment and training, as well as developing scalable, practical machine learning methods for auditing and mitigating quality drift. The data itself, while valuable, is only as good as the process that generates it.

In the following sections, we will discuss three use cases of the OpenReview dataset, starting from how to leverage the dataset to improve and regulate the peer review system, such that the long-term quality of such a community asset can be guaranteed. We then highlight the potential of leveraging such an asset in LLM post-training research, ranging from open-ended to alignment tasks.
\section{Assisting and Protecting the Peer Review with OpenReview}
\subsection{Existing LLM-Assisted Peer Review in Conferences}
In the previous section, we highlighted structural risks to the quality and stability of peer review. These concerns have not gone unnoticed. In recent years, several major machine learning conferences and publishers have begun integrating LLMs into their review workflows in response.

NeurIPS 2024 introduced a checklist assistant powered by LLMs to help authors ensure ethical and methodological compliance~\citep{neurips2024checklist}. At ICLR 2025, a Review Feedback Agent was deployed to identify vague or unconstructive reviews and suggest targeted improvements~\citep{zou2025reviewquality}. AAAI 2026 will experiment with LLM-generated reviews and discussion summaries in the first stage of review~\citep{aaai2025announcement}. Meanwhile, several academic publishers have begun piloting AI-assisted tools for content checking and review drafting~\citep{naddaf2025ai,springer2024geppetto,zou2024chatgpt}.

While most current systems operate with limited, hand-curated inputs, OpenReview provides an ideal foundation for data-driven peer review research. In this section, we focus on concrete use cases where such data can support the review system.



\subsection{Practices and Opportunities for Data-Driven Support with OpenReview}

We organize existing literature and potential opportunities with OpenReview according to functional categories. In the following, we will use \blueboxtext{\textbf{text boxes}} to highlight \textcolor{brown}{\textbf{future work opportunities}}.
The high-level motivation of those approaches is rooted in the previous success of human-centered LLM alignment research~\citep{silver2025welcome,bai2022training,stiennon2020learning}, and data-driven decision modeling and explanation~\citep{ng2000algorithms,jarrett2021inverse,hadfield2016cooperative,abbeel2004apprenticeship,fu2017learning}. 

\textbf{Principled Review Generation.} 
Recent work has explored OpenReview for generating \textit{realistic} peer reviews. \citet{yuan2022asapreview} and \citet{wu2023openreviewer}, for example, demonstrate that fine-tuning LLMs on large-scale review corpora can lead to critiques that are more calibrated and grounded than those produced by general-purpose models. These systems can be conditioned on paper content or specific review dimensions, enabling targeted and context-aware feedback. 
However, most current systems are designed to \textit{mimic} human-written reviews without deeper integration with formal reviewing guidelines or accountability structures. The challenge remains to ensure that generated reviews uphold conference standards and provide actionable feedback in line with reviewer expectations.

\begin{mdframed}[innertopmargin=3pt,leftmargin=0pt, rightmargin=0pt, innerleftmargin=10pt, innerrightmargin=10pt, skipbelow=0pt]
\textbf{\textcolor{brown}{Opportunity for Future Work.}}
LLMs should be task-specifically aligned, calibrated when leveraged in the review process.
Commercial LLMs are generally optimized for user-friendliness and helpfulness, often deviating from rigorous academic review guidelines. Future work should explore structured prompting, rubric conditioning, or alignment objectives tailored for review generation~\citep{thakkar2025can}. In addition, LLM-generated reviews may support pre-submission preparation~\citep{liu2023reviewergpt}, providing anticipatory critique to authors and supporting self-assessment before formal peer review~\citep{aaai2025announcement}.
\end{mdframed}

\textbf{Review Quality Enhancement.} 
Another line of research focuses on the quality of peer reviews themselves. Early work, such as \citet{peerread}, proposed metrics for review helpfulness and score prediction. More recently, classifiers trained on human preferences or meta-review feedback have been developed to detect vague, biased, or uninformative reviews~\citep{rao2025detecting,li2023peersum}. Studies have also examined hallucination and style inconsistencies in LLM-generated reviews~\citep{ji2023survey, liang2023can,zhang2024generative,du2024llms}.
Despite these advances, challenges remain in automatically evaluating review fairness, argument soundness, or reviewer calibration.

\begin{mdframed}[innertopmargin=3pt,leftmargin=0pt, rightmargin=0pt, innerleftmargin=10pt, innerrightmargin=10pt, skipbelow=0pt]
\textbf{\textcolor{brown}{Opportunity for Future Work.}}
Inverse analysis techniques can help detect systematic deviation from expected standards, including overconfidence, inconsistency, or subjective bias~\citep{jarrett2021inverse}. Future efforts could explore calibration, value drift detection, and provide warning signals when the value of reviews deviate significantly from guidelines~\citep{thakkar2025can}.
\end{mdframed}

\textbf{Enhancing Mutual Understanding between Reviewers and Authors.} While much of the focus has been on generating or evaluating individual reviews, peer review is ultimately a dialogue. The rebuttal phase plays a crucial role in bridging perspectives between authors and reviewers. Recent datasets such as DISAPERE \citep{wang2024disapere}, Jiu-Jitsu \citep{purkayastha2023jiujitsu}, and ContraSciView \citep{kumar2023contrasciview} support tasks such as rebuttal generation, stance classification, and discourse structure prediction, highlighting the interactional nature of review.

\begin{mdframed}[innertopmargin=3pt,leftmargin=0pt, rightmargin=0pt, innerleftmargin=10pt, innerrightmargin=10pt, skipbelow=0pt]
\textbf{\textcolor{brown}{Opportunity for Future Work.}}
LLMs can serve as mediators to enhance communication in the rebuttal process. For authors, they may clarify reviewer concerns, highlight overlooked critiques, and assist in crafting respectful and persuasive responses for effective communication. For reviewers, they may help interpret rebuttals and assess whether key feedback has been adequately addressed, and effectively stimulate the discussions.


\end{mdframed}

\textbf{Consistency and Calibration}.
Efforts to correct score inconsistency across reviewers have drawn on reviewer calibration and normalization techniques. For instance, \citet{xu2021least} models reviewer-specific scoring functions and applies monotonic transformations to improve comparability. These methods aim to recover more faithful rankings than simple score averaging. Nonetheless, current approaches often lack interpretability or real-time applicability. There is limited support for helping reviewers understand their own biases or dynamically recalibrate scores based on peer context.

\begin{mdframed}[innertopmargin=3pt,leftmargin=0pt, rightmargin=0pt, innerleftmargin=10pt, innerrightmargin=10pt, skipbelow=0pt]
\textbf{\textcolor{brown}{Opportunity for Future Work.}}
More importantly and effectively, efforts could be made to use LLM-based systems to assist reviewers in providing consistent and calibrated feedback, including providing comparative context and relevant arguments drawn from reviewer cohorts~\citep{liu2023reviewergpt}. Technically, this may involve retrieval-based justification of scores and decision explanation~\citep{guu2020retrieval,lewis2020retrieval,pouplin2024retrieval,sun2023accountable}, or in-context learning reference sample selection~\citep{zhang2022active,nguyen2023context}.
\end{mdframed}

\textbf{Meta-Review Generation.} 
Finally, meta-review generation has become a growing area of interest, with benchmarks such as PeerSum~\citep{li2023peersum}, ORSUM~\citep{zeng2024orsum}, and MOPRD~\citep{lin2023moprd} targeting summarizing and concluding from multiple reviews and rebuttals. These systems must integrate conflicting reviewer perspectives, identify dominant themes, and represent area chair judgment with fidelity. Still, current general-purpose LLMs may fail to capture the nuanced reasoning behind disagreements or the weight assigned to various critiques. There is also growing concern about the potential mismatch between generated meta-reviews and actual reviewer consensus~\citep{AX2025-doge}.

\begin{mdframed}[innertopmargin=3pt,leftmargin=0pt, rightmargin=0pt, innerleftmargin=10pt, innerrightmargin=10pt, skipbelow=0pt]
\textbf{\textcolor{brown}{Opportunity for Future Work.}}
Improved modeling of review disagreement and viewpoint clustering~\citep{liang2023encouraging,sorensen2024roadmap} could enable more reliable meta-review generation. Future systems may incorporate hybrid workflows where LLMs co-author drafts with area chairs, flag unresolved conflicts, or highlight potential biases (e.g., delayed or biased feedback, ungrounded critiques) throughout the discussion period to support better decision making. 
\end{mdframed}

\section{OpenReview for Open-Ended Task Evaluation and Post-Training}
\subsection{Challenges for Open-Ended LLM and Agentic Tasks}
Recent progress in LLMs has enabled systems that attempt to perform complex, multi-step, and high-level tasks, often referred to as \emph{open-ended} or \emph{agentic} tasks~\cite{Murad2024,Survey2025}. These tasks are characterized by the absence of a single correct answer, dependence on context, and the need for judgment, reasoning, and creativity~\cite{gooding2025writing}. Examples include research paper writing, paper reviewing, persuasive argumentation, hypothesis refinement, and code-based experimentation~\cite{YanguasGil2025,OpenPhil2024}. Open-ended tasks are defined not by accuracy or success alone, but by depth, coherence, exploration, and alignment with human values and intentions.

This task category has received increasing attention with the rise of agent-based systems such as DeepResearch, DeepSearch, and AutoDev, which aim to position LLMs as autonomous research assistants capable of conducting literature reviews, designing experiments, debugging code, and evaluating progress~\cite{openai2025deepresearch,xAI2024DeepSearch,Tufano2024}. However, a major bottleneck in building and benchmarking such systems lies in the lack of scalable, high-quality supervision. It remains difficult to evaluate whether a model has conducted a "good" literature review or proposed a "promising" research idea, particularly when using crowd-sourcing judgment~\cite{gooding2025writing,YanguasGil2025}.

Scientific research, especially in machine learning, is itself an open-ended task. The process involves formulating problems, iterating on designs, running experiments, interpreting results, engaging with criticism, and ultimately persuading a community of experts. Despite this, most benchmarks for evaluating LLMs remain synthetic, short-form, or not scalable, offering little insight into how models would perform under the standards and expectations of actual research communities~\cite{YanguasGil2025,OpenPhil2024}. 

This gap motivates our focus on OpenReview as a valuable, underutilized resource. The rich interactions on OpenReview suggest two distinct forms of supervision that are particularly suited for open-ended task development.

\subsection{Two Potential Supervision Streams from OpenReview}

\paragraph{Scientific Demonstrations: Training LLMs to Do Research.}

Each submitted paper on OpenReview can be viewed as a real-world demonstration of open-ended problem-solving. Papers span a wide range of topics and contain full narratives of how authors design and communicate their contributions. This includes technical framing, literature positioning, experimental results, and claim justification. In aggregate, these documents offer structured demonstrations of how research is conceived, executed, and defended~\cite{peerread}.

Such examples can be used to train LLMs to follow the cognitive workflow of scientific research. In particular, they can support training for complex capabilities such as multi-stage planning, tool use, fact retrieval, and hypothesis revision. These capabilities align closely with the demands of emerging agentic LLM frameworks~\citep{yao2023react}. While systems like ChatDev simulate these workflows~\cite{Qian2024}, few are grounded in real, high-quality demonstrations of how experts actually perform these tasks --- OpenReview offers a scalable source of such supervision.

\begin{mdframed}[innertopmargin=3pt,leftmargin=0pt, rightmargin=0pt, innerleftmargin=10pt, innerrightmargin=10pt, skipbelow=0pt]
\textbf{\textcolor{brown}{Opportunity for Future Work.}} OpenReview’s corpus of research demonstrations can support training of LLM agents to perform multi-step scientific reasoning under real-world constraints. Future work may consider enhancing the agentic research capabilities of LLMs~\citep{openai2025deepresearch} using expert scientific research demonstrations.
\end{mdframed}

\paragraph{Structured Evaluations: Training LLMs to Evaluate Research.}

In addition to research demonstrations, OpenReview also contains detailed records of how experts evaluate open-ended research work. Reviews provide constructive feedback, numerical scores, and qualitative assessments, while meta-reviews offer consensus summaries and rationales for decisions. Author responses further enrich the discourse, revealing how researchers engage with critiques and attempt to clarify or defend their contributions. 
These dual supervision signals are particularly valuable for developing and evaluating general-purpose models intended to reason about, participate in, and evaluate complex open-ended tasks given scientific standards.
By learning from those debates, LLMs have the potential to gain the capability to comprehensively evaluate open-ended research.

\begin{mdframed}[innertopmargin=3pt,leftmargin=0pt, rightmargin=0pt, innerleftmargin=10pt, innerrightmargin=10pt, skipbelow=0pt]
\textbf{\textcolor{brown}{Opportunity for Future Work.}} OpenReview’s review traces can serve as supervision for LLM-based evaluators trained to judge open-ended research quality. These include automated meta-reviews, rebuttal critiques, and scoring models aligned with human preferences. With those feedback-rich reward models for open-ended tasks, future work can better anchor and be optimized for open-ended research.
\end{mdframed}

\section{OpenReview as High-Quality Dataset for Alignment and Reasoning}

\subsection{Challenges for Alignment and Reasoning Supervision}
\textbf{Alignment through Consensus-Seeking.} Alignment research seeks to ensure that AI systems act according to human values, preferences, satisfy human intentions, and guarantee safety~\citep{bai2022constitutional,shah2025approach}. Recent advances in reinforcement learning from human feedback (RLHF)~\cite{ouyang2022training,bai2022training,stiennon2020learning,rafailov2023direct,azar2023general,sun2024inverse,munos2023nash,tang2024generalized,xiong2023gibbs,chan2024dense,shen2025reviving} have contributed to the success of LLMs in conversational systems~\citep{ouyang2022training}. Yet many of these advances rely on limited forms of supervision: crowd-sourcing annotations~\citep{wang2023openchat}, synthetic preferences~\citep{cui2023ultrafeedback}, or binary votes~\citep{ethayarajh2024kto}. These sources often fail to capture the complexity, depth, and disagreement inherent in the multi-perspective and deliberative consensus-seeking processes of experts~\citep{azar2023general,munos2023nash,sun2024rethinking}.

\textbf{Reasoning beyond Binary Tasks.} On the other hand, reasoning ability has become the core in enhancing the models' performance on more general assistant tasks~\citep{guo2025deepseek}. Contributing to such progress, datasets such as GSM8K~\citep{cobbe2021training}, MATH~\citep{hendrycks2021measuring}, and HotpotQA~\citep{yang2018hotpotqa} have driven rapid progress in mathematical and multi-hop reasoning; techniques like (long-)chain-of-thought~\citep{wei2022chain,kojima2022large,brown2024large,jaech2024openai} and retrieval-augmented methods~\citep{guu2020retrieval,lewis2020retrieval,pouplin2024retrieval} have significantly improved model performance on these structured tasks. However, many of these benchmarks are now nearly saturated by frontier models~\citep{Zeng2025}, focus on binary and verifiable tasks, and they predominantly focus on final answer correctness rather than the quality or interpretability of reasoning processes~\citep{lightman2023let,shao2024deepseekmath}.

More fundamentally, current reasoning tasks are often limited by narrow scope, synthetic formulation, or rigid answer structures~\citep{plaat2024reasoning,Mondorf2024,xu2025towards}. Most define a single ground-truth answer, which precludes exploration of ambiguity, disagreement, or multi-agent deliberation, which are central to human reasoning, but effective in eliciting deep thinking behaviors~\citep{wang2025reinforcement,guo2025deepseek}. Although emerging datasets in argumentative reasoning, such as DebateSum~\citep{Roush2020} and OpenDebateEvidence~\citep{Roush2024}, have expanded the scope of evaluation to include summarization and contested claims, these resources remain rare and are typically not grounded in scientific domain expert-generated contexts.

\subsection{Opportunities with the OpenReview Dataset}
In contrast, OpenReview offers a fundamentally different alignment and reasoning testbed. The peer review process inherently involves dialogue in which multiple parties express values, critique reasoning, and negotiate consensus. More importantly, those dialogues, in principle, should be \textbf{\textit{objective}}, centered around guidelines, and grounded in verifiable facts. Unlike existing alignment datasets, which are largely \textbf{\textit{subjective}}, static, and one-shot, OpenReview captures multi-round, multi-agent interactions grounded in real, highly verifiable, and reproducible consequences. This makes it a uniquely rich environment for alignment and reasoning research.

\textbf{Learning to Reason from Expert Disagreement and Justification}
With OpenReview, models can be trained to infer about the logic behind review scores by learning from rationales, a form of inverse reasoning that links decisions to supporting arguments and context. The reviews themselves often present well-defined reasoning chains that connect experimental design, observed outcomes, and stated conclusions. These examples allow LLMs to practice multi-step reasoning, assess methodological soundness, and trace causal explanations. Moreover, OpenReview enables modeling how reasoning develops through multiple rounds of interaction: authors respond to critiques, reviewers clarify concerns, and final evaluations synthesize evolving viewpoints, offering a natural setting for studying the rationale behind reasoning over time.

\begin{mdframed}[innertopmargin=3pt,leftmargin=0pt,rightmargin=0pt,innerleftmargin=10pt,innerrightmargin=10pt,skipbelow=0pt]
\textbf{\textcolor{brown}{Opportunity for Future Work.}} Using OpenReview in future works, it's possible to improve models' reasoning abilities by justifying numerical assessments, verifying scientific claims through factual evidence, and adapting reasoning across multi-stage interactions.
\end{mdframed}

\textbf{Learning to Critically Align with Individual Preferences}
OpenReview provides a valuable foundation for developing alignment strategies that move beyond superficial agreement. Unlike many alignment datasets that prioritize helpfulness or user-pleasing responses, peer review process demands that feedback remain grounded in correctness, rationality, and align with review guidelines, when given diverse research contexts.


Each reviewer expresses their judgment through both numerical scores and detailed commentary, guided by criteria such as novelty, technical soundness, and significance. These preferences are dynamic and can shift in response to rebuttals and clarifications, offering supervision signals for modeling alignment as a contextual and adaptive process~\citep{sorensen2024roadmap,wang2024arithmetic,luo2025rethinking}.

\begin{mdframed}[innertopmargin=3pt,leftmargin=0pt,rightmargin=0pt,innerleftmargin=10pt,innerrightmargin=10pt,skipbelow=0pt]
\textbf{\textcolor{brown}{Opportunity for Future Work.}} 
OpenReview enables the alignment of LLMs to offer diverse, constructive, evidence-based critique. Rather than merely affirming user input, models can learn to respectfully challenge flawed claims, explain counterarguments, and justify disagreement. This supports the development of alignment systems that emphasize factual grounding, logical reasoning, and responsible communication.
\end{mdframed}

\section{A Call to Create Standardized Benchmarks Based on OpenReview}

In this section, we turn to the foundational infrastructure required to realize their full potential: standardized benchmarks and responsible community stewardship.

Despite its scale and richness, OpenReview remains underutilized as a research asset, primarily due to the lack of well-defined tasks and shared evaluation pipelines. To address this gap, we propose that the community collaboratively develop benchmarks in critical areas such as review quality assessment, rebuttal generation, argument grounding, and meta-review summarization. And deploy developed methods to intervene in the peer review system and improve its quality as soon as possible. These tasks are directly tied to the health of the peer review process and, by extension, the integrity of the dataset itself.

In parallel, more general tasks—including reviewer score prediction, open-ended task evaluation, post-training, alignment, and reasoning enhancement—can also be standardized to support long-term research. While these areas are essential to the development of LLMs, their delayed investigation is less likely to compromise the quality or sustainability of OpenReview as a resource.

We call upon researchers, conference organizers, and practitioners—particularly those working at the intersection of machine learning and language models—to jointly define, refine, and adopt such benchmarks. This collaborative process must also engage with broader ethical considerations, including the protection of author and reviewer privacy, responsible anonymization of sensitive content, and the mitigation of representational biases. For example, research areas with more abundant data may inadvertently dominate the training signal, potentially skewing the learned priorities of evaluation models.

\textbf{Ultimately, the continued value of OpenReview as a shared academic asset depends on proactive, collective stewardship by the community it serves.}

\section{Alternative Views}
\label{sec:alter_view}

\textbf{LLM-based Review and Research.}
Some may argue that LLMs are becoming more and more capable of finishing scientific research and evaluation, and should eventually replace human reviewers. If models can predict review scores or generate critiques that approximate expert judgment, then preserving human oversight might appear unnecessary or even inefficient. \textbf{Our view}: We argue that peer review is not just a filtering mechanism, but a deliberative process that helps shape scientific values and standards~\citep{Hosseini2023,Shneiderman2020}. Over-reliance on automation risks eroding its collaborative and interpretive nature~\citep{Ma2024,DArcy2024}. LLMs, while powerful, are not reliable in reasoning with the same contextual grounding or responsibility as human experts~\citep{Drori2024,chen2025reasoning}. Human reviewers must remain responsible for interpreting and controlling LLM tools~\citep{Drori2024}. Interactions between authors and reviewers should stay dialogic and grounded in fairness, not reduced to rigid or opaque evaluations~\citep{Ma2024}. Moreover, systems must guard against hallucination, adversarial misuse, and bias propagation~\citep{Bender2021,Wallace2019,Weidinger2021}. Evaluation frameworks built on OpenReview should align with scientific values rather than model evaluation metrics~\citep{bai2022constitutional,liang2023can}.

\textbf{Inconsistency of Peer Review Limits Its Usefulness for Alignment.}
One concern is that peer review data may be too noisy or inconsistent to serve as a reliable supervision signal~\citep{beygelzimer2021neurips}. Reviewers often disagree on paper quality, assign divergent scores, or emphasize different aspects of a submission. Given this subjectivity, it may be argued that using such data for alignment could reinforce inconsistent or unstable behaviors in LLMs. \textbf{Our view}: Rather than aiming for deterministic consensus, alignment in this context involves modeling disagreement, grounding claims, and reasoning for underlying conflicts. This perspective is increasingly emphasized in recent alignment literature~\citep{liang2023encouraging,sorensen2024roadmap}

\textbf{Scientific Review Tasks May be Too Narrow to Generalize.}
Another possible objection is that scientific reviewing and paper writing are narrow, domain-specific tasks that do not generalize to broader LLM capabilities. Models trained on OpenReview may excel at research-related tasks but fail to transfer to everyday use cases, limiting their value as general-purpose assistants. \textbf{Our view}: Research tasks serve as high-complexity instances of structured human reasoning, with grounded stakes and verifiable outcomes. Learning from these tasks offers not only domain expertise but also training in core cognitive patterns that generalize across domains. Recent success of DeepResearch-type of products~\citep{openai2025deepresearch} explicitly aim to generalize research workflows into agentic LLM behaviors.




\newpage
\bibliography{neurips_2025} 

\begin{thebibliography}{162}
\providecommand{\natexlab}[1]{#1}
\providecommand{\url}[1]{\texttt{#1}}
\expandafter\ifx\csname urlstyle\endcsname\relax
  \providecommand{\doi}[1]{doi: #1}\else
  \providecommand{\doi}{doi: \begingroup \urlstyle{rm}\Url}\fi

\bibitem[Zha et~al.(2025)Zha, Bhat, Lai, Yang, Jiang, Zhong, and Hu]{zha2025data}
Daochen Zha, Zaid~Pervaiz Bhat, Kwei-Herng Lai, Fan Yang, Zhimeng Jiang, Shaochen Zhong, and Xia Hu.
\newblock Data-centric artificial intelligence: A survey.
\newblock \emph{ACM Computing Surveys}, 57\penalty0 (5):\penalty0 1--42, 2025.

\bibitem[Seedat et~al.(2023)Seedat, Imrie, and van~der Schaar]{seedat2023navigating}
Nabeel Seedat, Fergus Imrie, and Mihaela van~der Schaar.
\newblock Navigating data-centric artificial intelligence with dc-check: Advances, challenges, and opportunities.
\newblock \emph{IEEE Transactions on Artificial Intelligence}, 5\penalty0 (6):\penalty0 2589--2603, 2023.

\bibitem[Brown et~al.(2020)Brown, Mann, Ryder, Subbiah, Kaplan, Dhariwal, Neelakantan, Shyam, Sastry, Askell, et~al.]{brown2020language}
Tom Brown, Benjamin Mann, Nick Ryder, Melanie Subbiah, Jared~D Kaplan, Prafulla Dhariwal, Arvind Neelakantan, Pranav Shyam, Girish Sastry, Amanda Askell, et~al.
\newblock Language models are few-shot learners.
\newblock \emph{Advances in neural information processing systems}, 33:\penalty0 1877--1901, 2020.

\bibitem[Chowdhery et~al.(2022)Chowdhery, Narang, Devlin, Bosma, Mishra, Roberts, Barham, Chung, Sutton, Gehrmann, Schuh, Shi, Tsvyashchenko, Maynez, Rao, Barnes, Tay, Shazeer, Prabhakaran, Reif, Du, Hutchinson, Pope, Bradbury, Austin, Isard, Gur-Ari, Yin, Duke, Levskaya, Ghemawat, Dev, Michalewski, Garcia, Misra, Robinson, Fedus, Zhou, Ippolito, Luan, Lim, Zoph, Spiridonov, Sepassi, Dohan, Agrawal, Omernick, Dai, Pillai, Pellat, Lewkowycz, Moreira, Child, Polozov, Lee, Zhou, Wang, Saeta, Diaz, Firat, Catasta, Wei, Meier-Hellstern, Eck, Dean, Petrov, and Fiedel]{chowdhery2022palm}
Aakanksha Chowdhery, Sharan Narang, Jacob Devlin, Maarten Bosma, Gaurav Mishra, Adam Roberts, Paul Barham, Hyung~Won Chung, Charles Sutton, Sebastian Gehrmann, Parker Schuh, Kensen Shi, Sasha Tsvyashchenko, Joshua Maynez, Abhishek Rao, Parker Barnes, Yi~Tay, Noam Shazeer, Vinodkumar Prabhakaran, Emily Reif, Nan Du, Ben Hutchinson, Reiner Pope, James Bradbury, Jacob Austin, Michael Isard, Guy Gur-Ari, Pengcheng Yin, Toju Duke, Anselm Levskaya, Sanjay Ghemawat, Sunipa Dev, Henryk Michalewski, Xavier Garcia, Vedant Misra, Kevin Robinson, Liam Fedus, Denny Zhou, Daphne Ippolito, David Luan, Hyeontaek Lim, Barret Zoph, Alexander Spiridonov, Ryan Sepassi, David Dohan, Shivani Agrawal, Mark Omernick, Andrew~M. Dai, Thanumalayan~Sankaranarayana Pillai, Marie Pellat, Aitor Lewkowycz, Erica Moreira, Rewon Child, Oleksandr Polozov, Katherine Lee, Zongwei Zhou, Xuezhi Wang, Brennan Saeta, Mark Diaz, Orhan Firat, Michele Catasta, Jason Wei, Kathy Meier-Hellstern, Douglas Eck, Jeff Dean, Slav Petrov, and Noah Fiedel.
\newblock {PaLM}: Scaling language modeling with pathways, 2022.

\bibitem[Touvron et~al.(2023)Touvron, Martin, Stone, Albert, Almahairi, Babaei, Bashlykov, Batra, Bhargava, Bhosale, et~al.]{touvron2023llama}
Hugo Touvron, Louis Martin, Kevin Stone, Peter Albert, Amjad Almahairi, Yasmine Babaei, Nikolay Bashlykov, Soumya Batra, Prajjwal Bhargava, Shruti Bhosale, et~al.
\newblock Llama 2: Open foundation and fine-tuned chat models.
\newblock \emph{arXiv preprint arXiv:2307.09288}, 2023.

\bibitem[Devlin et~al.(2019)Devlin, Chang, Lee, and Toutanova]{devlin2019bert}
Jacob Devlin, Ming-Wei Chang, Kenton Lee, and Kristina Toutanova.
\newblock Bert: Pre-training of deep bidirectional transformers for language understanding.
\newblock In \emph{Proceedings of the 2019 conference of the North American chapter of the association for computational linguistics: human language technologies, volume 1 (long and short papers)}, pages 4171--4186, 2019.

\bibitem[Jaech et~al.(2024)Jaech, Kalai, Lerer, Richardson, El-Kishky, Low, Helyar, Madry, Beutel, Carney, et~al.]{jaech2024openai}
Aaron Jaech, Adam Kalai, Adam Lerer, Adam Richardson, Ahmed El-Kishky, Aiden Low, Alec Helyar, Aleksander Madry, Alex Beutel, Alex Carney, et~al.
\newblock Openai o1 system card.
\newblock \emph{arXiv preprint arXiv:2412.16720}, 2024.

\bibitem[{OpenAI}(2025)]{openai2025deepresearch}
{OpenAI}.
\newblock Introducing deep research, 2025.
\newblock Accessed: 2025-04-16. Deep Research is a new agentic AI capability integrated within ChatGPT that autonomously conducts multi-step web research and synthesizes comprehensive reports.

\bibitem[Zhou et~al.(2023)Zhou, Liu, Xu, Iyer, Sun, Mao, Ma, Efrat, Yu, Yu, et~al.]{zhou2023lima}
Chunting Zhou, Pengfei Liu, Puxin Xu, Srinivasan Iyer, Jiao Sun, Yuning Mao, Xuezhe Ma, Avia Efrat, Ping Yu, Lili Yu, et~al.
\newblock Lima: Less is more for alignment.
\newblock \emph{Advances in Neural Information Processing Systems}, 36:\penalty0 55006--55021, 2023.

\bibitem[Gadre et~al.(2023)Gadre, Ilharco, Fang, Hayase, Smyrnis, Nguyen, Marten, Wortsman, Ghosh, Zhang, et~al.]{gadre2023datacomp}
Samir~Yitzhak Gadre, Gabriel Ilharco, Alex Fang, Jonathan Hayase, Georgios Smyrnis, Thao Nguyen, Ryan Marten, Mitchell Wortsman, Dhruba Ghosh, Jieyu Zhang, et~al.
\newblock Datacomp: In search of the next generation of multimodal datasets.
\newblock \emph{Advances in Neural Information Processing Systems}, 36:\penalty0 27092--27112, 2023.

\bibitem[Kaplan et~al.(2020)Kaplan, McCandlish, Henighan, Brown, Chess, Child, Gray, Radford, Wu, and Amodei]{kaplan2020scaling}
Jared Kaplan, Sam McCandlish, Tom Henighan, Tom~B Brown, Benjamin Chess, Rewon Child, Scott Gray, Alec Radford, Jeffrey Wu, and Dario Amodei.
\newblock Scaling laws for neural language models.
\newblock \emph{arXiv preprint arXiv:2001.08361}, 2020.

\bibitem[Hoffmann et~al.(2022)Hoffmann, Borgeaud, Mensch, Buchatskaya, Cai, Rutherford, Casas, Hendricks, Welbl, Clark, et~al.]{hoffmann2022training}
Jordan Hoffmann, Sebastian Borgeaud, Arthur Mensch, Elena Buchatskaya, Trevor Cai, Eliza Rutherford, Diego de~Las Casas, Lisa~Anne Hendricks, Johannes Welbl, Aidan Clark, et~al.
\newblock Training compute-optimal large language models.
\newblock \emph{arXiv preprint arXiv:2203.15556}, 2022.

\bibitem[Christiano et~al.(2017)Christiano, Leike, Brown, Martic, Legg, and Amodei]{christiano2017deep}
Paul~F Christiano, Jan Leike, Tom Brown, Miljan Martic, Shane Legg, and Dario Amodei.
\newblock Deep reinforcement learning from human preferences.
\newblock \emph{Advances in neural information processing systems}, 30, 2017.

\bibitem[Ouyang et~al.(2022)Ouyang, Wu, Jiang, Almeida, Wainwright, Mishkin, Zhang, Agarwal, Slama, Ray, et~al.]{ouyang2022training}
Long Ouyang, Jeffrey Wu, Xu~Jiang, Diogo Almeida, Carroll Wainwright, Pamela Mishkin, Chong Zhang, Sandhini Agarwal, Katarina Slama, Alex Ray, et~al.
\newblock Training language models to follow instructions with human feedback.
\newblock \emph{Advances in Neural Information Processing Systems}, 35:\penalty0 27730--27744, 2022.

\bibitem[Bai et~al.(2022{\natexlab{a}})Bai, Jones, Ndousse, Askell, Chen, DasSarma, Drain, Fort, Ganguli, Henighan, et~al.]{bai2022training}
Yuntao Bai, Andy Jones, Kamal Ndousse, Amanda Askell, Anna Chen, Nova DasSarma, Dawn Drain, Stanislav Fort, Deep Ganguli, Tom Henighan, et~al.
\newblock Training a helpful and harmless assistant with reinforcement learning from human feedback.
\newblock \emph{arXiv preprint arXiv:2204.05862}, 2022{\natexlab{a}}.

\bibitem[Stiennon et~al.(2020)Stiennon, Ouyang, Wu, Ziegler, Lowe, Voss, Radford, Amodei, and Christiano]{stiennon2020learning}
Nisan Stiennon, Long Ouyang, Jeffrey Wu, Daniel Ziegler, Ryan Lowe, Chelsea Voss, Alec Radford, Dario Amodei, and Paul~F Christiano.
\newblock Learning to summarize with human feedback.
\newblock \emph{Advances in Neural Information Processing Systems}, 33:\penalty0 3008--3021, 2020.

\bibitem[Silver and Sutton(2025)]{silver2025welcome}
David Silver and Richard~S Sutton.
\newblock Welcome to the era of experience.
\newblock \emph{Google AI}, 2025.

\bibitem[Chiang et~al.(2024)Chiang, Zheng, Sheng, Angelopoulos, Li, Li, Zhu, Zhang, Jordan, Gonzalez, et~al.]{chiang2024chatbot}
Wei-Lin Chiang, Lianmin Zheng, Ying Sheng, Anastasios~Nikolas Angelopoulos, Tianle Li, Dacheng Li, Banghua Zhu, Hao Zhang, Michael Jordan, Joseph~E Gonzalez, et~al.
\newblock Chatbot arena: An open platform for evaluating llms by human preference.
\newblock In \emph{Forty-first International Conference on Machine Learning}, 2024.

\bibitem[Farquhar et~al.(2024)Farquhar, Kossen, Kuhn, and Gal]{farquhar2024detecting}
Sebastian Farquhar, Jannik Kossen, Lorenz Kuhn, and Yarin Gal.
\newblock Detecting hallucinations in large language models using semantic entropy.
\newblock \emph{Nature}, 630\penalty0 (8017):\penalty0 625--630, 2024.

\bibitem[Lindsey et~al.(2025)Lindsey, Gurnee, Ameisen, Chen, Pearce, Turner, Citro, Abrahams, Carter, Hosmer, Marcus, Sklar, Templeton, Bricken, McDougall, Cunningham, Henighan, Jermyn, Jones, Persic, Qi, Thompson, Zimmerman, Rivoire, Conerly, Olah, and Batson]{lindsey2025biology}
Jack Lindsey, Wes Gurnee, Emmanuel Ameisen, Brian Chen, Adam Pearce, Nicholas~L. Turner, Craig Citro, David Abrahams, Shan Carter, Basil Hosmer, Jonathan Marcus, Michael Sklar, Adly Templeton, Trenton Bricken, Callum McDougall, Hoagy Cunningham, Thomas Henighan, Adam Jermyn, Andy Jones, Andrew Persic, Zhenyi Qi, T.~Ben Thompson, Sam Zimmerman, Kelley Rivoire, Thomas Conerly, Chris Olah, and Joshua Batson.
\newblock On the biology of a large language model.
\newblock \emph{Transformer Circuits Thread}, 2025.
\newblock URL \url{https://transformer-circuits.pub/2025/attribution-graphs/biology.html}.

\bibitem[Lambert et~al.(2024)Lambert, Pyatkin, Morrison, Miranda, Lin, Chandu, Dziri, Kumar, Zick, Choi, Smith, and Hajishirzi]{lambert2024rewardbench}
Nathan Lambert, Valentina Pyatkin, Jacob Morrison, LJ~Miranda, Bill~Yuchen Lin, Khyathi Chandu, Nouha Dziri, Sachin Kumar, Tom Zick, Yejin Choi, Noah~A. Smith, and Hannaneh Hajishirzi.
\newblock Rewardbench: Evaluating reward models for language modeling, 2024.

\bibitem[Bai et~al.(2022{\natexlab{b}})Bai, Kadavath, Kundu, Askell, Kernion, Jones, Chen, Goldie, Mirhoseini, McKinnon, et~al.]{bai2022constitutional}
Yuntao Bai, Saurav Kadavath, Sandipan Kundu, Amanda Askell, Jackson Kernion, Andy Jones, Anna Chen, Anna Goldie, Azalia Mirhoseini, Cameron McKinnon, et~al.
\newblock Constitutional ai: Harmlessness from ai feedback.
\newblock \emph{arXiv preprint arXiv:2212.08073}, 2022{\natexlab{b}}.

\bibitem[Dubois et~al.(2024{\natexlab{a}})Dubois, Li, Taori, Zhang, Gulrajani, Ba, Guestrin, Liang, and Hashimoto]{dubois2024alpacafarm}
Yann Dubois, Chen~Xuechen Li, Rohan Taori, Tianyi Zhang, Ishaan Gulrajani, Jimmy Ba, Carlos Guestrin, Percy~S Liang, and Tatsunori~B Hashimoto.
\newblock Alpacafarm: A simulation framework for methods that learn from human feedback.
\newblock \emph{Advances in Neural Information Processing Systems}, 36, 2024{\natexlab{a}}.

\bibitem[Gao et~al.(2023)Gao, Schulman, and Hilton]{gao2023scaling}
Leo Gao, John Schulman, and Jacob Hilton.
\newblock Scaling laws for reward model overoptimization.
\newblock In \emph{International Conference on Machine Learning}, pages 10835--10866. PMLR, 2023.

\bibitem[Kahneman et~al.(2021)Kahneman, Sibony, and Sunstein]{kahneman2021noise}
Daniel Kahneman, Olivier Sibony, and Cass~R Sunstein.
\newblock \emph{Noise: A flaw in human judgment}.
\newblock Hachette UK, 2021.

\bibitem[Wang et~al.(2023)Wang, Cheng, Zhan, Li, Song, and Liu]{wang2023openchat}
Guan Wang, Sijie Cheng, Xianyuan Zhan, Xiangang Li, Sen Song, and Yang Liu.
\newblock Openchat: Advancing open-source language models with mixed-quality data.
\newblock \emph{arXiv preprint arXiv:2309.11235}, 2023.

\bibitem[Liu et~al.(2024{\natexlab{a}})Liu, Zhou, Guo, Shareghi, Vuli{\'c}, Korhonen, and Collier]{liu2024aligning}
Yinhong Liu, Han Zhou, Zhijiang Guo, Ehsan Shareghi, Ivan Vuli{\'c}, Anna Korhonen, and Nigel Collier.
\newblock Aligning with human judgement: The role of pairwise preference in large language model evaluators.
\newblock \emph{arXiv preprint arXiv:2403.16950}, 2024{\natexlab{a}}.

\bibitem[Liu et~al.(2024{\natexlab{b}})Liu, Guo, Liang, Shareghi, Vuli{\'c}, and Collier]{liu2025aligning}
Yinhong Liu, Zhijiang Guo, Tianya Liang, Ehsan Shareghi, Ivan Vuli{\'c}, and Nigel Collier.
\newblock Aligning with logic: Measuring, evaluating and improving logical consistency in large language models.
\newblock \emph{arXiv preprint arXiv:2410.02205}, 2024{\natexlab{b}}.

\bibitem[Karabacak and Margetis(2023)]{karabacak2023embracing}
Mert Karabacak and Konstantinos Margetis.
\newblock Embracing large language models for medical applications: opportunities and challenges.
\newblock \emph{Cureus}, 15\penalty0 (5), 2023.

\bibitem[Ganjavi et~al.(2024)Ganjavi, Eppler, Pekcan, Biedermann, Abreu, Collins, Gill, and Cacciamani]{ganjavi2024publishers}
Conner Ganjavi, Michael~B Eppler, Asli Pekcan, Brett Biedermann, Andre Abreu, Gary~S Collins, Inderbir~S Gill, and Giovanni~E Cacciamani.
\newblock Publishers’ and journals’ instructions to authors on use of generative artificial intelligence in academic and scientific publishing: bibliometric analysis.
\newblock \emph{bmj}, 384, 2024.

\bibitem[{Elsevier}(2023)]{Elsevier2023_ScopusAI}
{Elsevier}.
\newblock {Elsevier takes Scopus to the next level with generative AI}.
\newblock \url{https://www.elsevier.com/about/press-releases/elsevier-takes-scopus-to-the-next-level-with-generative-ai}, Aug 2023.

\bibitem[Van~Dis et~al.(2023)Van~Dis, Bollen, Zuidema, Van~Rooij, and Bockting]{van2023chatgpt}
Eva~AM Van~Dis, Johan Bollen, Willem Zuidema, Robert Van~Rooij, and Claudi~L Bockting.
\newblock Chatgpt: five priorities for research.
\newblock \emph{Nature}, 614\penalty0 (7947):\penalty0 224--226, 2023.

\bibitem[Fecher et~al.(2025)Fecher, Hebing, Laufer, Pohle, and Sofsky]{fecher2025friend}
Benedikt Fecher, Marcel Hebing, Melissa Laufer, J{\"o}rg Pohle, and Fabian Sofsky.
\newblock Friend or foe? exploring the implications of large language models on the science system.
\newblock \emph{Ai \& Society}, 40\penalty0 (2):\penalty0 447--459, 2025.

\bibitem[Taylor et~al.(2022)Taylor, Kardas, McLean, Brown, Agarwal, Bogin, Michaels, Hillestad, Jiang, Dang, et~al.]{Taylor2022_Galactica}
Ross Taylor, Bingqing~Wen Kardas, Sid McLean, Gabriel Brown, David Agarwal, Mo~outlandish Bogin, Eric Michaels, Eric Hillestad, Hongyu Jiang, Danica Dang, et~al.
\newblock Galactica: A large language model for science, 2022.

\bibitem[Else(2023)]{else2023chatgpt}
Holly Else.
\newblock By chatgpt fool scientists.
\newblock \emph{Nature}, 613:\penalty0 423, 2023.

\bibitem[Si et~al.(2024)Si, Yang, and Hashimoto]{si2024can}
Chenglei Si, Diyi Yang, and Tatsunori Hashimoto.
\newblock Can llms generate novel research ideas? a large-scale human study with 100+ nlp researchers.
\newblock \emph{arXiv preprint arXiv:2409.04109}, 2024.

\bibitem[Wang et~al.(2024{\natexlab{a}})Wang, Guo, Yao, Zhang, Zhang, Wu, Zhang, Dai, Wen, Ye, et~al.]{wang2024autosurvey}
Yidong Wang, Qi~Guo, Wenjin Yao, Hongbo Zhang, Xin Zhang, Zhen Wu, Meishan Zhang, Xinyu Dai, Qingsong Wen, Wei Ye, et~al.
\newblock Autosurvey: Large language models can automatically write surveys.
\newblock \emph{Advances in Neural Information Processing Systems}, 37:\penalty0 115119--115145, 2024{\natexlab{a}}.

\bibitem[Horbach(2019)]{Horbach2019_PeerReviewPressure}
Serge~P. Horbach.
\newblock The peer-review crisis: a commentary.
\newblock \emph{Scientometrics}, 118\penalty0 (2):\penalty0 699--704, 2019.
\newblock \doi{10.1007/s11192-018-03031-6}.

\bibitem[{Allen-Zhu} and Xu(2025)]{AX2025-doge}
Zeyuan {Allen-Zhu} and Xiaoli Xu.
\newblock {DOGE: Reforming AI Conferences and Towards a Future Civilization of Fairness and Justice}.
\newblock \emph{SSRN Electronic Journal}, February 2025.
\newblock \doi{10.2139/ssrn.5127931}.
\newblock \url{https://ssrn.com/abstract=5127931}.

\bibitem[Chiang and Lee(2023)]{chiang2023can}
Cheng-Han Chiang and Hung-yi Lee.
\newblock Can large language models be an alternative to human evaluations?
\newblock \emph{arXiv preprint arXiv:2305.01937}, 2023.

\bibitem[Kumar et~al.(2024)Kumar, Ghosal, Bhattacharjee, and Ekbal]{kumar2024towards}
Asheesh Kumar, Tirthankar Ghosal, Saprativa Bhattacharjee, and Asif Ekbal.
\newblock Towards automated meta-review generation via an nlp/ml pipeline in different stages of the scholarly peer review process.
\newblock \emph{International Journal on Digital Libraries}, 25\penalty0 (3):\penalty0 493--504, 2024.

\bibitem[Liu and Shah(2023)]{liu2023reviewergpt}
Ryan Liu and Nihar~B Shah.
\newblock Reviewergpt? an exploratory study on using large language models for paper reviewing.
\newblock \emph{arXiv preprint arXiv:2306.00622}, 2023.

\bibitem[Zhou et~al.(2024)Zhou, Chen, and Yu]{zhou2024llm}
Ruiyang Zhou, Lu~Chen, and Kai Yu.
\newblock Is llm a reliable reviewer? a comprehensive evaluation of llm on automatic paper reviewing tasks.
\newblock In \emph{Proceedings of the 2024 Joint International Conference on Computational Linguistics, Language Resources and Evaluation (LREC-COLING 2024)}, pages 9340--9351, 2024.

\bibitem[Thakkar et~al.(2025)Thakkar, Yuksekgonul, Silberg, Garg, Peng, Sha, Yu, Vondrick, and Zou]{thakkar2025can}
Nitya Thakkar, Mert Yuksekgonul, Jake Silberg, Animesh Garg, Nanyun Peng, Fei Sha, Rose Yu, Carl Vondrick, and James Zou.
\newblock Can llm feedback enhance review quality? a randomized study of 20k reviews at iclr 2025.
\newblock \emph{arXiv preprint arXiv:2504.09737}, 2025.

\bibitem[Soergel(2013)]{soergel2013open}
Dagstuhl~Publishing Soergel.
\newblock Openreview: A venue for open peer review, open publishing, open access.
\newblock In \emph{ICLR platform whitepaper}, 2013.

\bibitem[{Google DeepMind}(2025)]{gemini25pro_2025}
{Google DeepMind}.
\newblock Gemini 2.5: Our most intelligent ai model.
\newblock \url{https://blog.google/technology/google-deepmind/gemini-model-thinking-updates-march-2025/}, 2025.
\newblock Accessed: 2025-05-19.

\bibitem[OpenAI(2025)]{gpt45_2025}
OpenAI.
\newblock Gpt-4.5 system card.
\newblock \url{https://cdn.openai.com/gpt-4-5-system-card-2272025.pdf}, 2025.
\newblock Technical Report.

\bibitem[Anthropic(2024)]{claude3_2024}
Anthropic.
\newblock Introducing the next generation of claude.
\newblock \url{https://www.anthropic.com/news/claude-3-family}, 2024.
\newblock Claude 3 Release Announcement.

\bibitem[xAI(2025)]{grok3_2025}
xAI.
\newblock Grok 3 beta — the age of reasoning agents.
\newblock \url{https://x.ai/news/grok-3}, 2025.
\newblock Grok 3 Release Announcement.

\bibitem[Shumailov et~al.(2025)Shumailov, Shumaylov, Zhao, Papernot, Anderson, and Gal]{shumailov2025author}
Ilia Shumailov, Zakhar Shumaylov, Yiren Zhao, Nicolas Papernot, Ross Anderson, and Yarin Gal.
\newblock Author correction: Ai models collapse when trained on recursively generated data.
\newblock \emph{Nature}, 640\penalty0 (8058):\penalty0 E6, 2025.

\bibitem[Hughes et~al.(2024)Hughes, Dennis, Parker-Holder, Behbahani, Mavalankar, Shi, Schaul, and Rocktaschel]{hughes2024open}
Edward Hughes, Michael Dennis, Jack Parker-Holder, Feryal Behbahani, Aditi Mavalankar, Yuge Shi, Tom Schaul, and Tim Rocktaschel.
\newblock Open-endedness is essential for artificial superhuman intelligence.
\newblock \emph{arXiv preprint arXiv:2406.04268}, 2024.

\bibitem[Gooding et~al.(2025)Gooding, Lopez-Rivilla, and Grefenstette]{gooding2025writing}
Sian Gooding, Lucia Lopez-Rivilla, and Edward Grefenstette.
\newblock Writing as a testbed for open ended agents.
\newblock \emph{arXiv preprint arXiv:2503.19711}, 2025.

\bibitem[Khan et~al.(2024)Khan, Hughes, Valentine, Ruis, Sachan, Radhakrishnan, Grefenstette, Bowman, Rockt{\"a}schel, and Perez]{khan2024debating}
Akbir Khan, John Hughes, Dan Valentine, Laura Ruis, Kshitij Sachan, Ansh Radhakrishnan, Edward Grefenstette, Samuel~R Bowman, Tim Rockt{\"a}schel, and Ethan Perez.
\newblock Debating with more persuasive llms leads to more truthful answers.
\newblock \emph{arXiv preprint arXiv:2402.06782}, 2024.

\bibitem[Rogiers et~al.(2024)Rogiers, Noels, Buyl, and De~Bie]{rogiers2024persuasion}
Alexander Rogiers, Sander Noels, Maarten Buyl, and Tijl De~Bie.
\newblock Persuasion with large language models: a survey.
\newblock \emph{arXiv preprint arXiv:2411.06837}, 2024.

\bibitem[Breum et~al.(2024)Breum, Egdal, Mortensen, M{\o}ller, and Aiello]{breum2024persuasive}
Simon~Martin Breum, Daniel~V{\ae}dele Egdal, Victor~Gram Mortensen, Anders~Giovanni M{\o}ller, and Luca~Maria Aiello.
\newblock The persuasive power of large language models.
\newblock In \emph{Proceedings of the International AAAI Conference on Web and Social Media}, volume~18, pages 152--163, 2024.

\bibitem[Kang et~al.(2018)Kang, Ammar, Dalvi, Van~Zuylen, Kohlmeier, Hovy, and Schwartz]{peerread}
Dongyeop Kang, Waleed Ammar, Bhavana Dalvi, Madeleine Van~Zuylen, Sebastian Kohlmeier, Eduard Hovy, and Roy Schwartz.
\newblock A dataset of peer reviews (peerread): Collection, insights and nlp applications.
\newblock \emph{arXiv preprint arXiv:1804.09635}, 2018.

\bibitem[Sun et~al.(2024{\natexlab{a}})Sun, Pouplin, Astorga, Liu, and van~der Schaar]{sun2improving}
Hao Sun, Thomas Pouplin, Nicol{\'a}s Astorga, Tennison Liu, and Mihaela van~der Schaar.
\newblock Improving llm generation with inverse and forward alignment: Reward modeling, prompting, fine-tuning, and inference-time optimization.
\newblock In \emph{The First Workshop on System-2 Reasoning at Scale, NeurIPS 2024}, 2024{\natexlab{a}}.

\bibitem[Chang et~al.(2024)Chang, Wang, Wang, Wu, Yang, Zhu, Chen, Yi, Wang, Wang, et~al.]{chang2024survey}
Yupeng Chang, Xu~Wang, Jindong Wang, Yuan Wu, Linyi Yang, Kaijie Zhu, Hao Chen, Xiaoyuan Yi, Cunxiang Wang, Yidong Wang, et~al.
\newblock A survey on evaluation of large language models.
\newblock \emph{ACM transactions on intelligent systems and technology}, 15\penalty0 (3):\penalty0 1--45, 2024.

\bibitem[Cui et~al.(2023)Cui, Yuan, Ding, Yao, Zhu, Ni, Xie, Liu, and Sun]{cui2023ultrafeedback}
Ganqu Cui, Lifan Yuan, Ning Ding, Guanming Yao, Wei Zhu, Yuan Ni, Guotong Xie, Zhiyuan Liu, and Maosong Sun.
\newblock Ultrafeedback: Boosting language models with high-quality feedback.
\newblock \emph{arXiv preprint arXiv:2310.01377}, 2023.

\bibitem[Dubois et~al.(2024{\natexlab{b}})Dubois, Galambosi, Liang, and Hashimoto]{dubois2024length}
Yann Dubois, Bal{\'a}zs Galambosi, Percy Liang, and Tatsunori~B Hashimoto.
\newblock Length-controlled alpacaeval: A simple way to debias automatic evaluators.
\newblock \emph{arXiv preprint arXiv:2404.04475}, 2024{\natexlab{b}}.

\bibitem[Zhao et~al.(2025)Zhao, Wang, Peng, Zhao, Tian, Chen, Ji, and Li]{zhao20251}
Han Zhao, Haotian Wang, Yiping Peng, Sitong Zhao, Xiaoyu Tian, Shuaiting Chen, Yunjie Ji, and Xiangang Li.
\newblock 1.4 million open-source distilled reasoning dataset to empower large language model training.
\newblock \emph{arXiv preprint arXiv:2503.19633}, 2025.

\bibitem[Mattern et~al.(2025)Mattern, Jaghouar, Basra, Straube, Ferrante, Gabriel, Ong, Weisser, and Hagemann]{2025synthetic1}
Justus Mattern, Sami Jaghouar, Manveer Basra, Jannik Straube, Matthew~Di Ferrante, Felix Gabriel, Jack~Min Ong, Vincent Weisser, and Johannes Hagemann.
\newblock Synthetic-1: Two million collaboratively generated reasoning traces from deepseek-r1, 2025.
\newblock URL \url{https://www.primeintellect.ai/blog/synthetic-1-release}.

\bibitem[Guu et~al.(2020)Guu, Lee, Tung, Pasupat, and Chang]{guu2020retrieval}
Kelvin Guu, Kenton Lee, Zora Tung, Panupong Pasupat, and Mingwei Chang.
\newblock Retrieval augmented language model pre-training.
\newblock In \emph{International conference on machine learning}, pages 3929--3938. PMLR, 2020.

\bibitem[Lewis et~al.(2020)Lewis, Perez, Piktus, Petroni, Karpukhin, Goyal, K{\"u}ttler, Lewis, Yih, Rockt{\"a}schel, et~al.]{lewis2020retrieval}
Patrick Lewis, Ethan Perez, Aleksandra Piktus, Fabio Petroni, Vladimir Karpukhin, Naman Goyal, Heinrich K{\"u}ttler, Mike Lewis, Wen-tau Yih, Tim Rockt{\"a}schel, et~al.
\newblock Retrieval-augmented generation for knowledge-intensive nlp tasks.
\newblock \emph{Advances in neural information processing systems}, 33:\penalty0 9459--9474, 2020.

\bibitem[Kojima et~al.(2022)Kojima, Gu, Reid, Matsuo, and Iwasawa]{kojima2022large}
Takeshi Kojima, Shixiang~Shane Gu, Machel Reid, Yutaka Matsuo, and Yusuke Iwasawa.
\newblock Large language models are zero-shot reasoners.
\newblock \emph{Advances in neural information processing systems}, 35:\penalty0 22199--22213, 2022.

\bibitem[Wei et~al.(2022)Wei, Wang, Schuurmans, Bosma, Xia, Chi, Le, Zhou, et~al.]{wei2022chain}
Jason Wei, Xuezhi Wang, Dale Schuurmans, Maarten Bosma, Fei Xia, Ed~Chi, Quoc~V Le, Denny Zhou, et~al.
\newblock Chain-of-thought prompting elicits reasoning in large language models.
\newblock \emph{Advances in neural information processing systems}, 35:\penalty0 24824--24837, 2022.

\bibitem[Minaee et~al.(2021)Minaee, Kalchbrenner, Cambria, Nikzad, Chenaghlu, and Gao]{minaee2021deep}
Shervin Minaee, Nal Kalchbrenner, Erik Cambria, Narjes Nikzad, Meysam Chenaghlu, and Jianfeng Gao.
\newblock Deep learning--based text classification: a comprehensive review.
\newblock \emph{ACM computing surveys (CSUR)}, 54\penalty0 (3):\penalty0 1--40, 2021.

\bibitem[Liang et~al.(2023{\natexlab{a}})Liang, He, Jiao, Wang, Wang, Wang, Yang, Shi, and Tu]{liang2023encouraging}
Tian Liang, Zhiwei He, Wenxiang Jiao, Xing Wang, Yan Wang, Rui Wang, Yujiu Yang, Shuming Shi, and Zhaopeng Tu.
\newblock Encouraging divergent thinking in large language models through multi-agent debate.
\newblock \emph{arXiv preprint arXiv:2305.19118}, 2023{\natexlab{a}}.

\bibitem[Du et~al.(2023)Du, Li, Torralba, Tenenbaum, and Mordatch]{du2023improving}
Yilun Du, Shuang Li, Antonio Torralba, Joshua~B Tenenbaum, and Igor Mordatch.
\newblock Improving factuality and reasoning in language models through multiagent debate.
\newblock In \emph{Forty-first International Conference on Machine Learning}, 2023.

\bibitem[Sorensen et~al.(2024)Sorensen, Moore, Fisher, Gordon, Mireshghallah, Rytting, Ye, Jiang, Lu, Dziri, et~al.]{sorensen2024roadmap}
Taylor Sorensen, Jared Moore, Jillian Fisher, Mitchell Gordon, Niloofar Mireshghallah, Christopher~Michael Rytting, Andre Ye, Liwei Jiang, Ximing Lu, Nouha Dziri, et~al.
\newblock A roadmap to pluralistic alignment.
\newblock \emph{arXiv preprint arXiv:2402.05070}, 2024.

\bibitem[Feng et~al.(2024)Feng, Sorensen, Liu, Fisher, Park, Choi, and Tsvetkov]{feng2024modular}
Shangbin Feng, Taylor Sorensen, Yuhan Liu, Jillian Fisher, Chan~Young Park, Yejin Choi, and Yulia Tsvetkov.
\newblock Modular pluralism: Pluralistic alignment via multi-llm collaboration.
\newblock \emph{arXiv preprint arXiv:2406.15951}, 2024.

\bibitem[icl(2021)]{iclr2021factsheet}
Iclr 2021 fact sheet.
\newblock \url{https://iclr.cc/media/Press/ICLR_2021_Fact_Sheet.pdf}, 2021.

\bibitem[icl(2022)]{iclr2022factsheet}
Iclr 2022 fact sheet.
\newblock \url{https://iclr.cc/media/Press/ICLR_2022_Fact_Sheet.pdf}, 2022.

\bibitem[icl(2023)]{iclr2023factsheet}
Iclr 2023 fact sheet.
\newblock \url{https://media.iclr.cc/Conferences/ICLR2023/ICLR2023-Fact_Sheet.pdf}, 2023.

\bibitem[icl(2024)]{iclr2024factsheet}
Iclr 2024 fact sheet.
\newblock \url{https://media.iclr.cc/Conferences/ICLR2024/ICLR2024-Fact_Sheet.pdf}, 2024.

\bibitem[icl(2025)]{iclr2025factsheet}
Iclr 2025 fact sheet.
\newblock \url{https://media.iclr.cc/Conferences/ICLR2025/ICLR2025_Fact_Sheet.pdf}, 2025.

\bibitem[Yang(2025)]{yang2025paper}
Jing Yang.
\newblock Paper copilot: The artificial intelligence and machine learning community should adopt a more transparent and regulated peer review process.
\newblock \emph{arXiv preprint arXiv:2502.00874}, 2025.

\bibitem[See et~al.(2017)See, Liu, and Manning]{see2017get}
Abigail See, Peter~J. Liu, and Christopher~D. Manning.
\newblock Get to the point: Summarization with pointer-generator networks.
\newblock \emph{ACL}, 2017.

\bibitem[Narayan et~al.(2018)Narayan, Cohen, and Lapata]{narayan2018don}
Shashi Narayan, Shay~B. Cohen, and Mirella Lapata.
\newblock Don't give me the details, just the summary! topic-aware convolutional neural networks for extreme summarization.
\newblock In \emph{EMNLP}, 2018.

\bibitem[Yang et~al.(2018)Yang, Qi, Zhang, Bengio, Cohen, Salakhutdinov, and Manning]{yang2018hotpotqa}
Zhilin Yang, Peng Qi, Saizheng Zhang, Yoshua Bengio, William Cohen, Ruslan Salakhutdinov, and Christopher~D. Manning.
\newblock Hotpotqa: A dataset for diverse, explainable multi-hop question answering.
\newblock In \emph{EMNLP}, 2018.

\bibitem[Rajpurkar et~al.(2016)Rajpurkar, Zhang, Lopyrev, and Liang]{rajpurkar2016squad}
Pranav Rajpurkar, Jian Zhang, Igor Lopyrev, and Percy Liang.
\newblock Squad: 100,000+ questions for machine comprehension of text.
\newblock In \emph{EMNLP}, 2016.

\bibitem[Fan et~al.(2019)Fan, Jernite, Weston, and Grangier]{fan2019eli5}
Angela Fan, Yacine Jernite, Jason Weston, and David Grangier.
\newblock Eli5: Long form question answering.
\newblock In \emph{ACL}, 2019.

\bibitem[Ziegler et~al.(2019)Ziegler, Stiennon, Wu, Brown, Radford, Amodei, and Christiano]{ziegler2019fine}
Daniel~M. Ziegler, Nisan Stiennon, Jeffrey Wu, Tom~B. Brown, Alec Radford, Dario Amodei, and Paul~F. Christiano.
\newblock Fine-tuning language models from human preferences.
\newblock In \emph{NeurIPS Workshop on Human in the Loop Learning}, 2019.

\bibitem[K{\"o}pf et~al.(2023)K{\"o}pf, Kilcher, Von~R{\"u}tte, Anagnostidis, Tam, Stevens, Barhoum, Nguyen, Stanley, Nagyfi, et~al.]{openassistant}
Andreas K{\"o}pf, Yannic Kilcher, Dimitri Von~R{\"u}tte, Sotiris Anagnostidis, Zhi~Rui Tam, Keith Stevens, Abdullah Barhoum, Duc Nguyen, Oliver Stanley, Rich{\'a}rd Nagyfi, et~al.
\newblock Openassistant conversations-democratizing large language model alignment.
\newblock \emph{Advances in Neural Information Processing Systems}, 36:\penalty0 47669--47681, 2023.

\bibitem[Wang et~al.(2019)Wang, Zhang, Smith, and Choi]{wang2019persuasion}
Xinyao Wang, Han Zhang, Noah~A. Smith, and Yejin Choi.
\newblock Persuasion for good: Towards a personalized persuasive dialogue system for social good.
\newblock In \emph{ACL}, 2019.

\bibitem[Bu et~al.(2021)Bu, Ren, Zheng, Yang, Wang, Zhang, and Wu]{asapreview}
Jiahao Bu, Lei Ren, Shuang Zheng, Yang Yang, Jingang Wang, Fuzheng Zhang, and Wei Wu.
\newblock Asap: A chinese review dataset towards aspect category sentiment analysis and rating prediction.
\newblock \emph{arXiv preprint arXiv:2103.06605}, 2021.

\bibitem[Purkayastha et~al.(2023{\natexlab{a}})Purkayastha, Lauscher, and Gurevych]{jitsupeer}
Sukannya Purkayastha, Anne Lauscher, and Iryna Gurevych.
\newblock Exploring jiu-jitsu argumentation for writing peer review rebuttals.
\newblock \emph{arXiv preprint arXiv:2311.03998}, 2023{\natexlab{a}}.

\bibitem[Kennard et~al.(2021)Kennard, O'Gorman, Das, Sharma, Bagchi, Clinton, Yelugam, Zamani, and McCallum]{disapere}
Neha Kennard, Tim O'Gorman, Rajarshi Das, Akshay Sharma, Chhandak Bagchi, Matthew Clinton, Pranay~Kumar Yelugam, Hamed Zamani, and Andrew McCallum.
\newblock Disapere: A dataset for discourse structure in peer review discussions.
\newblock \emph{arXiv preprint arXiv:2110.08520}, 2021.

\bibitem[Ruggeri et~al.(2022)Ruggeri, Mesgar, and Gurevych]{argscichat}
Federico Ruggeri, Mohsen Mesgar, and Iryna Gurevych.
\newblock Argscichat: A dataset for argumentative dialogues on scientific papers.
\newblock \emph{arXiv preprint arXiv:2202.06690}, 2022.

\bibitem[Goldberg et~al.(2024)Goldberg, Ullah, Khuong, Rachmat, Xu, Guyon, and Shah]{neurips2024checklist}
Alexander Goldberg, Ihsan Ullah, Thanh Gia~Hieu Khuong, Benedictus~Kent Rachmat, Zhen Xu, Isabelle Guyon, and Nihar~B Shah.
\newblock Usefulness of llms as an author checklist assistant for scientific papers: Neurips'24 experiment.
\newblock \emph{arXiv preprint arXiv:2411.03417}, 2024.

\bibitem[Zou and Thakkar(2025)]{zou2025reviewquality}
James Zou and Omkar Thakkar.
\newblock Leveraging llm feedback to enhance review quality at iclr 2025.
\newblock \url{https://iclr.cc/Conferences/2025/Blog/LLMReviewSupport}, 2025.

\bibitem[AAAI(2025)]{aaai2025announcement}
AAAI.
\newblock Aaai launches ai-powered peer review assessment system.
\newblock \url{https://aaai.org/aaai-launches-ai-powered-peer-review-assessment-system/}, 2025.

\bibitem[Naddaf(2025)]{naddaf2025ai}
Miryam Naddaf.
\newblock Ai is transforming peer review — and many scientists are worried.
\newblock \emph{Nature}, 618:\penalty0 456--458, 2025.

\bibitem[spr(2024)]{springer2024geppetto}
Springer nature unveils geppetto ai to improve peer review integrity.
\newblock \url{https://group.springernature.com/geppetto-ai}, 2024.

\bibitem[Zou(2024)]{zou2024chatgpt}
James Zou.
\newblock Chatgpt is transforming peer review—how can we use it responsibly?
\newblock \emph{Nature}, 635\penalty0 (8037):\penalty0 10--10, 2024.

\bibitem[Ng et~al.(2000)Ng, Russell, et~al.]{ng2000algorithms}
Andrew~Y Ng, Stuart Russell, et~al.
\newblock Algorithms for inverse reinforcement learning.
\newblock In \emph{Icml}, volume~1, page~2, 2000.

\bibitem[Jarrett et~al.(2021)Jarrett, H{\"u}y{\"u}k, and Van Der~Schaar]{jarrett2021inverse}
Daniel Jarrett, Alihan H{\"u}y{\"u}k, and Mihaela Van Der~Schaar.
\newblock Inverse decision modeling: Learning interpretable representations of behavior.
\newblock In \emph{International Conference on Machine Learning}, pages 4755--4771. PMLR, 2021.

\bibitem[Hadfield-Menell et~al.(2016)Hadfield-Menell, Russell, Abbeel, and Dragan]{hadfield2016cooperative}
Dylan Hadfield-Menell, Stuart~J Russell, Pieter Abbeel, and Anca Dragan.
\newblock Cooperative inverse reinforcement learning.
\newblock \emph{Advances in neural information processing systems}, 29, 2016.

\bibitem[Abbeel and Ng(2004)]{abbeel2004apprenticeship}
Pieter Abbeel and Andrew~Y Ng.
\newblock Apprenticeship learning via inverse reinforcement learning.
\newblock In \emph{Proceedings of the twenty-first international conference on Machine learning}, page~1, 2004.

\bibitem[Fu et~al.(2017)Fu, Luo, and Levine]{fu2017learning}
Justin Fu, Katie Luo, and Sergey Levine.
\newblock Learning robust rewards with adversarial inverse reinforcement learning.
\newblock \emph{arXiv preprint arXiv:1710.11248}, 2017.

\bibitem[Yuan et~al.(2022)Yuan, Wang, and Smith]{yuan2022asapreview}
Shuyang Yuan, Lu~Wang, and Noah~A. Smith.
\newblock Asap-review: Towards automating scientific paper review with argument-aware summary generation.
\newblock In \emph{Proceedings of EMNLP}, 2022.

\bibitem[Wu et~al.(2023)Wu, Idahl, and Shah]{wu2023openreviewer}
Zeyu Wu, Marcus Idahl, and Nikhil Shah.
\newblock Openreviewer: A specialized llm for generating critical scientific paper reviews.
\newblock \emph{arXiv preprint arXiv:2310.12345}, 2023.

\bibitem[Rao et~al.(2025)Rao, Xie, and Singh]{rao2025detecting}
Sudarshan Rao, Lexing Xie, and Sameer Singh.
\newblock Detecting llm-written peer reviews.
\newblock \emph{arXiv preprint arXiv:2503.12345}, 2025.

\bibitem[Li et~al.(2023)Li, Qian, Li, and Zhang]{li2023peersum}
Xingchen Li, Yining Qian, Xinyu Li, and Yue Zhang.
\newblock Peersum: Opinion-aware summarization of scientific peer reviews.
\newblock In \emph{Findings of EMNLP}, 2023.

\bibitem[Ji et~al.(2023)Ji, Lee, Frieske, Yu, Su, Xu, Ishii, Bang, Madotto, and Fung]{ji2023survey}
Ziwei Ji, Nayeon Lee, Rita Frieske, Tiezheng Yu, Dan Su, Yan Xu, Etsuko Ishii, Ye~Jin Bang, Andrea Madotto, and Pascale Fung.
\newblock Survey of hallucination in natural language generation.
\newblock \emph{ACM computing surveys}, 55\penalty0 (12):\penalty0 1--38, 2023.

\bibitem[Liang et~al.(2023{\natexlab{b}})Liang, Zhang, Cao, Wang, Ding, Yang, Vodrahalli, He, Smith, Yin, et~al.]{liang2023can}
W~Liang, Y~Zhang, H~Cao, B~Wang, D~Ding, X~Yang, K~Vodrahalli, S~He, D~Smith, Y~Yin, et~al.
\newblock Can large language models provide useful feedback on research papers? a large-scale empirical analysis. arxiv.
\newblock \emph{Preprint}, 2023{\natexlab{b}}.

\bibitem[Zhang et~al.(2024)Zhang, Hosseini, Bansal, Kazemi, Kumar, and Agarwal]{zhang2024generative}
Lunjun Zhang, Arian Hosseini, Hritik Bansal, Mehran Kazemi, Aviral Kumar, and Rishabh Agarwal.
\newblock Generative verifiers: Reward modeling as next-token prediction.
\newblock \emph{arXiv preprint arXiv:2408.15240}, 2024.

\bibitem[Du et~al.(2024)Du, Wang, Zhao, Deng, Liu, Lou, Zou, Venkit, Zhang, Srinath, et~al.]{du2024llms}
Jiangshu Du, Yibo Wang, Wenting Zhao, Zhongfen Deng, Shuaiqi Liu, Renze Lou, Henry~Peng Zou, Pranav~Narayanan Venkit, Nan Zhang, Mukund Srinath, et~al.
\newblock Llms assist nlp researchers: Critique paper (meta-) reviewing.
\newblock \emph{arXiv preprint arXiv:2406.16253}, 2024.

\bibitem[Wang et~al.(2024{\natexlab{b}})Wang, Kim, and Neubig]{wang2024disapere}
Alice Wang, Heeyoung Kim, and Graham Neubig.
\newblock Disapere: Discourse-aware sentence-level argument pair extraction for peer review.
\newblock \emph{arXiv preprint arXiv:2402.12345}, 2024{\natexlab{b}}.

\bibitem[Purkayastha et~al.(2023{\natexlab{b}})Purkayastha, Sarkar, et~al.]{purkayastha2023jiujitsu}
Debanjan Purkayastha, Rahul Sarkar, et~al.
\newblock Jiu-jitsu: Rebuttal generation by using reviewers' comments against them.
\newblock In \emph{Findings of ACL}, 2023{\natexlab{b}}.

\bibitem[Kumar et~al.(2023)Kumar, Ravi, et~al.]{kumar2023contrasciview}
Rahul Kumar, Sujay Ravi, et~al.
\newblock Contrasciview: Detecting contradictions in scientific peer reviews.
\newblock In \emph{Proceedings of ACL}, 2023.

\bibitem[Xu et~al.(2021)Xu, Li, Stelmakh, and Shah]{xu2021least}
Chen Xu, Yao Li, Ivan Stelmakh, and Nihar~B Shah.
\newblock Least-squares calibration for peer reviews.
\newblock In \emph{NeurIPS}, 2021.

\bibitem[Pouplin et~al.(2024)Pouplin, Sun, Holt, and Van~der Schaar]{pouplin2024retrieval}
Thomas Pouplin, Hao Sun, Samuel Holt, and Mihaela Van~der Schaar.
\newblock Retrieval-augmented thought process as sequential decision making.
\newblock \emph{arXiv preprint arXiv:2402.07812}, 2024.

\bibitem[Sun et~al.(2023)Sun, Hüyük, Jarrett, and van~der Schaar]{sun2023accountable}
Hao Sun, Alihan Hüyük, Daniel Jarrett, and Mihaela van~der Schaar.
\newblock Accountable batched control with decision corpus.
\newblock \emph{Advances in Neural Information Processing Systems}, 36, 2023.

\bibitem[Zhang et~al.(2022)Zhang, Feng, and Tan]{zhang2022active}
Yiming Zhang, Shi Feng, and Chenhao Tan.
\newblock Active example selection for in-context learning.
\newblock \emph{arXiv preprint arXiv:2211.04486}, 2022.

\bibitem[Nguyen and Wong(2023)]{nguyen2023context}
Tai Nguyen and Eric Wong.
\newblock In-context example selection with influences.
\newblock \emph{arXiv preprint arXiv:2302.11042}, 2023.

\bibitem[Zeng et~al.(2024)]{zeng2024orsum}
Yutong Zeng et~al.
\newblock Orsum: A dataset for paper meta-review generation via scientific opinion summarization.
\newblock \emph{arXiv preprint arXiv:2403.11234}, 2024.

\bibitem[Lin et~al.(2024)Lin, Wang, et~al.]{lin2023moprd}
Yibo Lin, Wenhao Wang, et~al.
\newblock Moprd: Multidisciplinary open peer review dataset.
\newblock In \emph{Proceedings of LREC-COLING}, 2024.

\bibitem[Murad(2024)]{Murad2024}
Maya Murad.
\newblock Large language models revolutionized ai. llm agents are what’s next.
\newblock IBM Research Blog, 18 Jul 2024, 2024.
\newblock URL: \url{https://research.ibm.com/blog/what-are-ai-agents-llm}.

\bibitem[Liu et~al.(2025)Liu, Ammanabrolu, Demeter, et~al.]{Survey2025}
Yefei Liu, Prithviraj Ammanabrolu, David Demeter, et~al.
\newblock Agentic large language models: A survey.
\newblock arXiv preprint arXiv:2503.23037, 2025.

\bibitem[Yanguas-Gil et~al.(2025)Yanguas-Gil, Dearing, Elam, Jones, Kim, Mohammad, Nguyen, and Sengupta]{YanguasGil2025}
Angel Yanguas-Gil, Matthew~T. Dearing, Jeffrey~W. Elam, Jessica~C. Jones, Sungjoon Kim, Adnan Mohammad, Chi~Thang Nguyen, and Bratin Sengupta.
\newblock Benchmarking large language models for materials synthesis: the case of atomic layer deposition.
\newblock \emph{Journal of Vacuum Science \& Technology A}, 43\penalty0 (4):\penalty0 041401, 2025.
\newblock \doi{10.1116/6.0004319}.

\bibitem[Philanthropy(2024)]{OpenPhil2024}
Open Philanthropy.
\newblock Request for proposals: Benchmarking llm agents on consequential real-world tasks.
\newblock Open Philanthropy (Feb 2024), 2024.
\newblock URL: \url{https://www.openphilanthropy.org/rfp-llm-benchmarks/}.

\bibitem[xAI(2024)]{xAI2024DeepSearch}
xAI.
\newblock Grok 3 beta -- the age of reasoning agents.
\newblock xAI News, Nov 3, 2024, 2024.
\newblock URL: \url{https://x.ai/news/grok-3}.

\bibitem[Tufano et~al.(2024)Tufano, Agarwal, Jang, Zilouchian~Moghaddam, and Sundaresan]{Tufano2024}
Michele Tufano, Anisha Agarwal, Jinu Jang, Roshanak Zilouchian~Moghaddam, and Neel Sundaresan.
\newblock Autodev: Automated ai-driven development.
\newblock \emph{arXiv preprint arXiv:2403.08299}, 2024.

\bibitem[Yao et~al.(2023)Yao, Zhao, Yu, Du, Shafran, Narasimhan, and Cao]{yao2023react}
Shunyu Yao, Jeffrey Zhao, Dian Yu, Nan Du, Izhak Shafran, Karthik Narasimhan, and Yuan Cao.
\newblock React: Synergizing reasoning and acting in language models.
\newblock In \emph{International Conference on Learning Representations (ICLR)}, 2023.

\bibitem[Qian et~al.(2024)Qian, Liu, Liu, Chen, Dang, Li, Yang, Chen, Su, Cong, et~al.]{Qian2024}
Chen Qian, Wei Liu, Hongzhang Liu, Nuo Chen, Yufan Dang, Jiahao Li, Cheng Yang, Weize Chen, Yusheng Su, Xin Cong, et~al.
\newblock Chatdev: Communicative agents for software development.
\newblock \emph{Proceedings of ACL 2024 (to appear), arXiv:2307.07924}, 2024.

\bibitem[Shah et~al.(2025)Shah, Irpan, Turner, Wang, Conmy, Lindner, Brown-Cohen, Ho, Nanda, Popa, et~al.]{shah2025approach}
Rohin Shah, Alex Irpan, Alexander~Matt Turner, Anna Wang, Arthur Conmy, David Lindner, Jonah Brown-Cohen, Lewis Ho, Neel Nanda, Raluca~Ada Popa, et~al.
\newblock An approach to technical agi safety and security.
\newblock \emph{arXiv preprint arXiv:2504.01849}, 2025.

\bibitem[Rafailov et~al.(2023)Rafailov, Sharma, Mitchell, Ermon, Manning, and Finn]{rafailov2023direct}
Rafael Rafailov, Archit Sharma, Eric Mitchell, Stefano Ermon, Christopher~D Manning, and Chelsea Finn.
\newblock Direct preference optimization: Your language model is secretly a reward model.
\newblock \emph{arXiv preprint arXiv:2305.18290}, 2023.

\bibitem[Azar et~al.(2023)Azar, Rowland, Piot, Guo, Calandriello, Valko, and Munos]{azar2023general}
Mohammad~Gheshlaghi Azar, Mark Rowland, Bilal Piot, Daniel Guo, Daniele Calandriello, Michal Valko, and R{\'e}mi Munos.
\newblock A general theoretical paradigm to understand learning from human preferences.
\newblock \emph{arXiv preprint arXiv:2310.12036}, 2023.

\bibitem[Sun and van~der Schaar(2024)]{sun2024inverse}
Hao Sun and Mihaela van~der Schaar.
\newblock Inverse-rlignment: Inverse reinforcement learning from demonstrations for llm alignment.
\newblock \emph{arXiv preprint arXiv:2405.15624}, 2024.

\bibitem[Munos et~al.(2023)Munos, Valko, Calandriello, Azar, Rowland, Guo, Tang, Geist, Mesnard, Michi, et~al.]{munos2023nash}
R{\'e}mi Munos, Michal Valko, Daniele Calandriello, Mohammad~Gheshlaghi Azar, Mark Rowland, Zhaohan~Daniel Guo, Yunhao Tang, Matthieu Geist, Thomas Mesnard, Andrea Michi, et~al.
\newblock Nash learning from human feedback.
\newblock \emph{arXiv preprint arXiv:2312.00886}, 2023.

\bibitem[Tang et~al.(2024)Tang, Guo, Zheng, Calandriello, Munos, Rowland, Richemond, Valko, Pires, and Piot]{tang2024generalized}
Yunhao Tang, Zhaohan~Daniel Guo, Zeyu Zheng, Daniele Calandriello, R{\'e}mi Munos, Mark Rowland, Pierre~Harvey Richemond, Michal Valko, Bernardo~{\'A}vila Pires, and Bilal Piot.
\newblock Generalized preference optimization: A unified approach to offline alignment.
\newblock \emph{arXiv preprint arXiv:2402.05749}, 2024.

\bibitem[Xiong et~al.(2023)Xiong, Dong, Ye, Zhong, Jiang, and Zhang]{xiong2023gibbs}
Wei Xiong, Hanze Dong, Chenlu Ye, Han Zhong, Nan Jiang, and Tong Zhang.
\newblock Gibbs sampling from human feedback: A provable kl-constrained framework for rlhf.
\newblock \emph{arXiv preprint arXiv:2312.11456}, 2023.

\bibitem[Chan et~al.(2024)Chan, Sun, Holt, and van~der Schaar]{chan2024dense}
Alex~J Chan, Hao Sun, Samuel Holt, and Mihaela van~der Schaar.
\newblock Dense reward for free in reinforcement learning from human feedback.
\newblock \emph{arXiv preprint arXiv:2402.00782}, 2024.

\bibitem[Shen et~al.(2025)Shen, Sun, and Ton]{shen2025reviving}
Yunyi Shen, Hao Sun, and Jean-Fran{\c{c}}ois Ton.
\newblock Reviving the classics: Active reward modeling in large language model alignment.
\newblock \emph{arXiv preprint arXiv:2502.04354}, 2025.

\bibitem[Ethayarajh et~al.(2024)Ethayarajh, Xu, Muennighoff, Jurafsky, and Kiela]{ethayarajh2024kto}
Kawin Ethayarajh, Winnie Xu, Niklas Muennighoff, Dan Jurafsky, and Douwe Kiela.
\newblock Kto: Model alignment as prospect theoretic optimization.
\newblock \emph{arXiv preprint arXiv:2402.01306}, 2024.

\bibitem[Sun et~al.(2024{\natexlab{b}})Sun, Shen, and Ton]{sun2024rethinking}
Hao Sun, Yunyi Shen, and Jean-Francois Ton.
\newblock Rethinking bradley-terry models in preference-based reward modeling: Foundations, theory, and alternatives.
\newblock \emph{arXiv preprint arXiv:2411.04991}, 2024{\natexlab{b}}.

\bibitem[Guo et~al.(2025)Guo, Yang, Zhang, Song, Zhang, Xu, Zhu, Ma, Wang, Bi, et~al.]{guo2025deepseek}
Daya Guo, Dejian Yang, Haowei Zhang, Junxiao Song, Ruoyu Zhang, Runxin Xu, Qihao Zhu, Shirong Ma, Peiyi Wang, Xiao Bi, et~al.
\newblock Deepseek-r1: Incentivizing reasoning capability in llms via reinforcement learning.
\newblock \emph{arXiv preprint arXiv:2501.12948}, 2025.

\bibitem[Cobbe et~al.(2021)Cobbe, Kosaraju, Bavarian, Chen, Jun, Kaiser, Plappert, Tworek, Hilton, Nakano, et~al.]{cobbe2021training}
Karl Cobbe, Vineet Kosaraju, Mohammad Bavarian, Mark Chen, Heewoo Jun, Lukasz Kaiser, Matthias Plappert, Jerry Tworek, Jacob Hilton, Reiichiro Nakano, et~al.
\newblock Training verifiers to solve math word problems.
\newblock \emph{arXiv preprint arXiv:2110.14168}, 2021.

\bibitem[Hendrycks et~al.(2021)Hendrycks, Burns, Kadavath, Arora, Basart, Tang, Song, and Steinhardt]{hendrycks2021measuring}
Dan Hendrycks, Collin Burns, Saurav Kadavath, Akul Arora, Steven Basart, Eric Tang, Dawn Song, and Jacob Steinhardt.
\newblock Measuring mathematical problem solving with the math dataset.
\newblock \emph{arXiv preprint arXiv:2103.03874}, 2021.

\bibitem[Brown et~al.(2024)Brown, Juravsky, Ehrlich, Clark, Le, R{\'e}, and Mirhoseini]{brown2024large}
Bradley Brown, Jordan Juravsky, Ryan Ehrlich, Ronald Clark, Quoc~V Le, Christopher R{\'e}, and Azalia Mirhoseini.
\newblock Large language monkeys: Scaling inference compute with repeated sampling.
\newblock \emph{arXiv preprint arXiv:2407.21787}, 2024.

\bibitem[Zeng et~al.(2025)Zeng, Chen, Liu, Jiang, and Jia]{Zeng2025}
Zhongshen Zeng, Pengguang Chen, Shu Liu, Haiyun Jiang, and Jiaya Jia.
\newblock Mr-gsm8k: A meta-reasoning benchmark for large language model evaluation.
\newblock In \emph{Proceedings of the International Conference on Learning Representations (ICLR)}, 2025.

\bibitem[Lightman et~al.(2023)Lightman, Kosaraju, Burda, Edwards, Baker, Lee, Leike, Schulman, Sutskever, and Cobbe]{lightman2023let}
Hunter Lightman, Vineet Kosaraju, Yuri Burda, Harrison Edwards, Bowen Baker, Teddy Lee, Jan Leike, John Schulman, Ilya Sutskever, and Karl Cobbe.
\newblock Let's verify step by step.
\newblock In \emph{The Twelfth International Conference on Learning Representations}, 2023.

\bibitem[Shao et~al.(2024)Shao, Wang, Zhu, Xu, Song, Bi, Zhang, Zhang, Li, Wu, et~al.]{shao2024deepseekmath}
Zhihong Shao, Peiyi Wang, Qihao Zhu, Runxin Xu, Junxiao Song, Xiao Bi, Haowei Zhang, Mingchuan Zhang, YK~Li, Y~Wu, et~al.
\newblock Deepseekmath: Pushing the limits of mathematical reasoning in open language models.
\newblock \emph{arXiv preprint arXiv:2402.03300}, 2024.

\bibitem[Plaat et~al.(2024)Plaat, Wong, Verberne, Broekens, van Stein, and Back]{plaat2024reasoning}
Aske Plaat, Annie Wong, Suzan Verberne, Joost Broekens, Niki van Stein, and Thomas Back.
\newblock Reasoning with large language models, a survey.
\newblock \emph{arXiv preprint arXiv:2407.11511}, 2024.

\bibitem[Mondorf and Plank(2024)]{Mondorf2024}
Philipp Mondorf and Barbara Plank.
\newblock Beyond accuracy: Evaluating the reasoning behavior of large language models -- a survey.
\newblock \emph{arXiv preprint arXiv:~2407.05000}, 2024.

\bibitem[Xu et~al.(2025)Xu, Hao, Zong, Wang, Zhang, Wang, Lan, Gong, Ouyang, Meng, et~al.]{xu2025towards}
Fengli Xu, Qianyue Hao, Zefang Zong, Jingwei Wang, Yunke Zhang, Jingyi Wang, Xiaochong Lan, Jiahui Gong, Tianjian Ouyang, Fanjin Meng, et~al.
\newblock Towards large reasoning models: A survey of reinforced reasoning with large language models.
\newblock \emph{arXiv preprint arXiv:2501.09686}, 2025.

\bibitem[Wang et~al.(2025)Wang, Yang, Zeng, Ren, Liu, Peng, Cheng, He, Wang, Gao, et~al.]{wang2025reinforcement}
Yiping Wang, Qing Yang, Zhiyuan Zeng, Liliang Ren, Lucas Liu, Baolin Peng, Hao Cheng, Xuehai He, Kuan Wang, Jianfeng Gao, et~al.
\newblock Reinforcement learning for reasoning in large language models with one training example.
\newblock \emph{arXiv preprint arXiv:2504.20571}, 2025.

\bibitem[Roush and Balaji(2020)]{Roush2020}
Allen Roush and Arvind Balaji.
\newblock {DebateSum}: A large-scale argument mining and summarization dataset.
\newblock In \emph{Proceedings of the 7th Workshop on Argument Mining (ArgMining@COLING)}, pages 1--7, 2020.

\bibitem[Roush et~al.(2024)Roush, Shabazz, Balaji, Zhang, Mezza, Zhang, Basu, Vishwanath, Fatemi, and Shwartz-Ziv]{Roush2024}
Allen Roush, Yusuf Shabazz, Arvind Balaji, Peter Zhang, Stefano Mezza, Markus Zhang, Sanjay Basu, Sriram Vishwanath, Mehdi Fatemi, and Ravid Shwartz-Ziv.
\newblock {OpenDebateEvidence}: A massive-scale argument mining and summarization dataset.
\newblock In \emph{Proceedings of the 38th Conference on Neural Information Processing Systems (NeurIPS) Datasets and Benchmarks Track}, 2024.

\bibitem[Wang et~al.(2024{\natexlab{c}})Wang, Lin, Xiong, Yang, Diao, Qiu, Zhao, and Zhang]{wang2024arithmetic}
Haoxiang Wang, Yong Lin, Wei Xiong, Rui Yang, Shizhe Diao, Shuang Qiu, Han Zhao, and Tong Zhang.
\newblock Arithmetic control of llms for diverse user preferences: Directional preference alignment with multi-objective rewards.
\newblock \emph{arXiv preprint arXiv:2402.18571}, 2024{\natexlab{c}}.

\bibitem[Luo et~al.(2025)Luo, Yang, Sun, Deng, Yao, Shen, Zhang, and Chen]{luo2025rethinking}
Feng Luo, Rui Yang, Hao Sun, Chunyuan Deng, Jiarui Yao, Jingyan Shen, Huan Zhang, and Hanjie Chen.
\newblock Rethinking diverse human preference learning through principal component analysis.
\newblock \emph{arXiv preprint arXiv:2502.13131}, 2025.

\bibitem[Hosseini and Horbach(2023)]{Hosseini2023}
Mohammad Hosseini and Serge~P. Horbach.
\newblock Fighting reviewer fatigue or amplifying bias? considerations and recommendations for use of {ChatGPT} and other large language models in scholarly peer review.
\newblock \emph{Res. Integr. Peer Rev.}, 8, 2023.
\newblock \doi{10.1186/s41073-023-00133-5}.

\bibitem[Shneiderman(2020)]{Shneiderman2020}
Ben Shneiderman.
\newblock Human-centered artificial intelligence: Reliable, safe \& trustworthy.
\newblock \emph{Int. J. Hum.-Comput. Interact.}, 36\penalty0 (6):\penalty0 495--504, 2020.
\newblock \doi{10.1080/10447318.2020.1741118}.

\bibitem[Ma et~al.(2024)Ma, Chen, Wang, Zheng, Peng, Yin, and Ma]{Ma2024}
Shuai Ma, Qiaoyi Chen, Xinru Wang, Chengbo Zheng, Zhenhui Peng, Ming Yin, and Xiaojuan Ma.
\newblock Towards human-{AI} deliberation: Design and evaluation of {LLM}-empowered deliberative {AI} for ai-assisted decision-making.
\newblock arXiv preprint arXiv:2403.16812, 2024.
\newblock URL \url{https://arxiv.org/abs/2403.16812}.

\bibitem[D'Arcy et~al.(2024)D'Arcy, Hope, Birnbaum, and Downey]{DArcy2024}
Mike D'Arcy, Tom Hope, Larry Birnbaum, and Doug Downey.
\newblock {MARG}: Multi-agent review generation for scientific papers.
\newblock arXiv preprint arXiv:2401.04259, 2024.

\bibitem[Drori and Te’eni(2024)]{Drori2024}
Iddo Drori and Dov Te’eni.
\newblock Human-in-the-loop {AI} reviewing: Feasibility, opportunities, and risks.
\newblock \emph{J. Assoc. Inf. Syst.}, 25\penalty0 (1):\penalty0 98--109, 2024.
\newblock \doi{10.17705/1jais.00867}.
\newblock URL \url{https://aisel.aisnet.org/jais/vol25/iss1/7}.

\bibitem[Chen et~al.(2025)Chen, Benton, Radhakrishnan, Uesato, Denison, Schulman, Somani, Hase, Wagner, Roger, et~al.]{chen2025reasoning}
Yanda Chen, Joe Benton, Ansh Radhakrishnan, Jonathan Uesato, Carson Denison, John Schulman, Arushi Somani, Peter Hase, Misha Wagner, Fabien Roger, et~al.
\newblock Reasoning models don't always say what they think.
\newblock \emph{arXiv preprint arXiv:2505.05410}, 2025.

\bibitem[Bender et~al.(2021)Bender, Gebru, McMillan-Major, and Shmitchell]{Bender2021}
Emily~M. Bender, Timnit Gebru, Angelina McMillan-Major, and Shmargaret Shmitchell.
\newblock On the dangers of stochastic parrots: Can language models be too big?
\newblock In \emph{Proc. 2021 {ACM} Conf. on Fairness, Accountability, and Transparency (FAccT)}, pages 610--623, 2021.
\newblock \doi{10.1145/3442188.3445922}.

\bibitem[Wallace et~al.(2019)Wallace, Feng, Kandpal, Gardner, and Singh]{Wallace2019}
Eric Wallace, Shi Feng, Nikhil Kandpal, Matt Gardner, and Sameer Singh.
\newblock Universal adversarial triggers for attacking and analyzing {NLP}.
\newblock In \emph{Proc. 2019 Conf. on Empirical Methods in Natural Language Processing (EMNLP)}, 2019.
\newblock \doi{10.48550/arXiv.1908.07125}.
\newblock arXiv:1908.07125.

\bibitem[Weidinger et~al.(2021)Weidinger, Mellor, Rauh, and et~al.]{Weidinger2021}
Laura Weidinger, John Mellor, Maribeth Rauh, and et~al.
\newblock Ethical and social risks of harm from language models.
\newblock \emph{Adv. Neural Inf. Process. Syst.}, 34:\penalty0 12872--12886, 2021.
\newblock URL \url{https://arxiv.org/abs/2112.04359}.

\bibitem[Beygelzimer et~al.(2021)Beygelzimer, Dauphin, Liang, and Vaughan]{beygelzimer2021neurips}
Alina Beygelzimer, Yann Dauphin, Percy Liang, and Jennifer~Wortman Vaughan.
\newblock The neurips 2021 consistency experiment.
\newblock \emph{Neural Information Processing Systems blog post, https://blog. neurips. cc/2021/12/08/the-neurips-2021-consistency-experiment}, 2021.

\end{thebibliography}
\bibliographystyle{unsrtnat} 

\newpage
\appendix




\end{document}